\documentclass[12pt,oneside,onecolumn,a4paper]{IEEEtran}
\usepackage{amsmath,epsfig,bbm}
\usepackage{amsmath}
\usepackage{gensymb}
\usepackage{graphicx}
\usepackage{color}
\usepackage[ruled]{algorithm2e}
\usepackage{multirow}

\def\comment#1{}

\newcommand{\fff}{\mathbbm{f}}

\newcommand{\bh}{{\bf h}}

\newcommand{\FS}{\mathbbm{F}_\mathcal{S}}

\begin{document}

\title{Spatial Source Subtraction Based on Incomplete Measurements of Relative Transfer Function}

\author{{\bf Zbyn\v{e}k
Koldovsk\'{y}$^a$, Ji\v{r}\'i M\'alek$^a$, and Sharon Gannot$^b$}
\vspace{0.1in} \\
$^a$Faculty of Mechatronics, Informatics, and Interdisciplinary
Studies, \\Technical University of Liberec,\\ Studentsk\'a 2, 461 17
Liberec, Czech Republic.\\ E-mail:
\{zbynek.koldovsky, jiri.malek\}@tul.cz,\\ fax:+420-485-353112, tel:+420-485-353534\\
$^b$Faculty of Engineering,
Bar-Ilan University,
Ramat-Gan, 5290002,
Israel.\\ E-mail: Sharon.Gannot@biu.ac.il, fax: 972-3-7384051, tel: 972-3-531-7618
}


\maketitle

\footnotetext{This work was supported by The Czech Sciences Foundation through Project No. 14-11898S.}

\begin{abstract}
Relative impulse responses between microphones are usually long and dense due to the reverberant acoustic environment. Estimating them from short and noisy recordings poses a long-standing challenge of audio signal processing. In this paper we apply a novel strategy based on ideas of Compressed Sensing. Relative transfer function (RTF) corresponding to the relative impulse response can often be estimated accurately from noisy data but only for certain frequencies. This means that often only an incomplete measurement of the RTF is available. A complete RTF estimate can be obtained through finding its sparsest representation in the time-domain: that is, through computing the sparsest among the corresponding relative impulse responses. Based on this approach, we propose to estimate the RTF from noisy data in three steps. First, the RTF is estimated using any conventional method such as the non-stationarity-based estimator by Gannot et al. or through Blind Source Separation. Second, frequencies are determined for which the RTF estimate appears to be accurate. Third, the RTF is reconstructed through solving a weighted $\ell_1$ convex program, which we propose to solve via a computationally efficient variant of the SpaRSA (Sparse Reconstruction by Separable Approximation) algorithm. An extensive experimental study with real-world recordings has been conducted. It has been shown that the proposed method is capable of improving many conventional estimators used as the first step in most situations.
\end{abstract}

{\keywords Relative Transfer Function, Relative Impulse Response, Sparse Approximations, $\ell_1$ norm, Compressed Sensing}

\section{Introduction}
Noise reduction, speech enhancement and signal separation have been goals in audio signal processing for decades. Although various methods were already proposed and also applied in practice, there still remain open problems. The main reason is that the propagation of sound in a natural acoustic environment is complex. Acoustical signals are wideband in nature and span a frequency range from 20~Hz to 20~kHz. Typical room impulse responses have thousands of coefficients; this aspect makes them difficult to estimate, especially in noisy conditions.

When dealing with, e.g., noise reduction, the crucial question is ``what is the unwanted part of the signal to be removed?'' Single-channel methods, most of which were developed earlier than multichannel methods, typically rely on some knowledge of noise or interference spectra. For example, the spectra can be acquired during noise-only periods, provided that information about the target source activity is available; see an overview of single-channel methods, e.g., in \cite{loizou,tashev,gannotcohen}. Multichannel methods can also use  spatial information \cite{gannotcohen,enhancement}. For example, a multichannel filter can be designed to cancel the signal coming from the target's position.  The output of this filter contains only noise and interference components and provides the key reference for the signal enhancement tasks.

Several terms are used in connection with the target signal cancelation, in particular, spatial source subtraction, null beamforming, target cancelation filter, and blocking matrix (BM). The latter refers to one of the building blocks of the minimum variance distortionless (MVDR) beamformer implemented in a generalized sidelobe canceler structure \cite{GSC}. The BM block is responsible for steering a null towards the desired source, hence blocking, yielding noise-only reference signals, further used to enhance the desired source  
through adaptive interference canceler and/or by a postfilter.  
Null beamformers were originally designed under the assumption of free-field propagation (no reverberation) knowing the microphone array geometry (e.g. linear or circular). But later they were also designed taking the reverberation into account; see, e.g., \cite{gannot,affes}. 

In natural acoustic environments, the reverberation must be taken into account to achieve satisfactory signal cancelation. This could be done knowing relative impulse responses or, equivalently, relative transfer functions (RTFs) between microphones \cite{gannot}. The RTF depends on the properties of the environment and on the positions of the target source and microphones. It can be easily computed from noise-free recordings when the target is static \cite{krueger, doclo}. However, the environment as well as the position of the target source can change quickly. Therefore, methods capable of estimating current RTF within short intervals of noisy recordings, during which the target is approximately static, are desirable.

There have been many attempts to estimate the RTF, or (more generally speaking) to design a null beamformer, from noisy recordings \cite{gannot,shmulik,zhao}. A popular approach is to use Blind Source Separation (BSS) based on Independent Component Analysis (ICA). However, the accuracy of ICA declines with the number of estimated parameters as it is a statistical approach \cite{cardoso}. The blind estimation of the RTF thus poses a challenging problem since there are thousands of coefficients (parameters) to be estimated. The difficulty of this task particularly grows with growing reverberation time and with growing distance of the target source.  A recent goal has therefore been to simplify the task through incorporation of prior knowledge. For example, the knowledge of approximate direction-of-arrival of the target is used in \cite{francesco,kellermann}, or a set of pre-estimated RTFs for potential positions of the target is assumed in \cite{icassp2013,semiblind,gannottalmon}. 

A novel strategy is used in \cite{lin,osher,icassp2014a} by considering the fact that relative impulse responses can be replaced or approximated by sparse filters, that is, by filters that have many coefficients equal to zero; see also \cite{benichoux}, a recent work on sparse approximations of room impulse responses. The authors of \cite{icassp2014a} propose a semi-blind approach assuming knowledge of the support of a sparse approximation. Hence only nonzero coefficients are estimated using ICA, which implies a significant dimensionality reduction of the parameter space. Results show that sparse estimates of filters achieve better target cancelation than dense filters that are estimated in a fully blind way. However, the assumption that the filter support is known is rather impractical.

In this paper, we propose a novel method based on the idea that the RTF could be known or accurately estimated only in several frequency bins. An appropriate name for such observation is {\em the incomplete measurement of the RTF}. The entire RTF is then reconstructed by finding a sparse representation of the incomplete measurement in the time-domain. In other words, the relative impulse response between the microphones is replaced by a sparse impulse response whose Fourier transform is, for known frequencies, (approximately) equal to the incomplete RTF. In fact, the idea draws on Compressed Sensing usually applied to sparse/compressible signals or images \cite{CS} as well as to system identification.

The following Section introduces the audio mixture model. Section III describes several methods to estimate the relative impulse response or the RTF, both when noise is or is not active. Section IV describes the proposed method, in which the incomplete RTF is reconstructed by an algorithm solving a weighted LASSO program with $\ell_1$ sparsity-inducing regularization. Section V then describes several ways to select the incomplete RTF estimate. Section VI presents an extensive experimental study with real recordings, and Section VII concludes this article.

\section{Problem Description}
\subsection{Model}
We will consider situations where two microphones are available\footnote{In this paper, we focus only on the two-microphone scenario due to its comparatively easy accessibility. The idea, however, may be generalized to more microphones. }. 
A stereo noisy observation of a target signal $s(n)$ can be described as
\begin{equation}\label{modeltime}
	\begin{split}
		x_{\rm L}(n)&=\{h_{\rm L}\ast s\}(n)+y_{\rm L}(n)\\
		x_{\rm R}(n)&=\{h_{\rm R}\ast s\}(n)+y_{\rm R}(n)
	\end{split}
\end{equation}
where $n$ is the time index taking values $1,\dots,N$; $\ast$ denotes the convolution; $x_{\rm L}$ and $x_{\rm R}$ are, respectively, the signals from the left and right microphones; and $y_{\rm L}$ and $y_{\rm R}$ are the remaining signals (noise and interferences) commonly referred to as noise. Further, $h_{\rm L}$ and $h_{\rm R}$ denote the microphone-target acoustical impulse responses. 
The signals as well as the impulse responses are supposed to be real-valued. 

This model assumes that the position of the target source remains (approximately) fixed during the recording interval, i.e., for $n=1,\dots,N$.

Using the relative impulse response between the microphones denoted as $h_{\rm rel}$, (\ref{modeltime}) can be re-written as
\begin{equation}\label{modeltime2}
\begin{split}
	x_{\rm L}(n)&=s_{\rm L}(n)+y_{\rm L}(n)\\
	x_{\rm R}(n)&=\{h_{\rm rel}\ast s_{\rm L}\}(n)+y_{\rm R}(n)
\end{split}
\end{equation}
where	$s_{\rm L}(n)=\{h_{\rm L}\ast s\}(n)$ and $h_{\rm rel}=h_{\rm L}^{-1}\ast h_{\rm R}$ where 
$h_{\rm L}^{-1}$ denotes the filter inverse  to $h_{\rm L}$. Note that although real-world acoustic channels $h_{\rm L}$ and $h_{\rm R}$ are causal, $h_{\rm rel}$ need not be so.

The equivalent description of (\ref{modeltime}) in the short-term frequency-domain is
\begin{equation}\label{modelfrekv}
\begin{split}
	X_{\rm L}(\theta,\ell)&=H_{\rm L}(\theta)S(\theta,\ell) + Y_{\rm L}(\theta,\ell),\\
	X_{\rm R}(\theta,\ell)&=H_{\rm R}(\theta)S(\theta,\ell) + Y_{\rm R}(\theta,\ell),
\end{split}
\end{equation}
where $\theta$ denotes the frequency, and $\ell$ is the frame index. 
The analogy to (\ref{modeltime2}) is
\begin{equation}\label{modelfrekv2}
\begin{split}
	X_{\rm L}(\theta,\ell)&=S_{\rm L}(\theta,\ell)+Y_{\rm L}(\theta,\ell)\\
	X_{\rm R}(\theta,\ell)&=H_{\rm RTF}(\theta) S_{\rm L}(\theta,\ell)+Y_{\rm R}(\theta,\ell)
\end{split}
\end{equation}
where $S_{\rm L}(\theta,\ell)=H_{\rm L}(\theta)S(\theta,\ell)$. Here $H_{\rm RTF}(\theta)$ denotes the Fourier transform of $h_{\rm rel}$, which is called the relative transfer function (RTF). It holds that
\[
	H_{\rm RTF}(\theta)=\frac{H_{\rm R}(\theta)}{H_{\rm L}(\theta)}.
\]
With low impact on generality, we assume that $H_{\rm L}$ does not have any zeros on the unit circle; see the discussion in \cite{gannot} on page~1619.

\subsection{Spatial Subtraction of a Target Source}\label{noisesection}
When $h_{\rm rel}$ or $H_{\rm RTF}$ are known, an efficient multichannel filter can be designed that cancels the target signal and only pass through noise signals. 
Consider two-input single-output filter defined as such that its output is 
\begin{equation}\label{noiseest1}
z=h\ast x_{\rm L}-x_{\rm R}.
\end{equation}
According to (\ref{modeltime2}), it holds that
\begin{equation}\label{bm}
z=\underbrace{(h-h_{\rm rel})\ast s_{\rm L}}_\text{target signal leakage} + \underbrace{h\ast y_{\rm L}-y_{\rm R}}_\text{noise reference}.
\end{equation}
For $h=h_{\rm rel}$, the target signal leakage vanishes, and
\begin{equation}\label{noiseest2}
z=h_{\rm rel}\ast y_{\rm L}-y_{\rm R}.
\end{equation}
This is the information provided about the noise signals $y_{\rm L}$ and $y_{\rm R}$, which is crucial in signal separation/enhancement or noise reduction applications. For example, the filter defined through (\ref{noiseest1}) serves as the blocking matrix part in systems having the structure of generalized sidelobe canceler, see, e.g., \cite{gannot, krueger, doclo, hoshuyama, takahashi,schwarz}. 

To complete the enhancement of the noisy signal, many steps still have to be taken, all of which pose other problems. For example, the spectrum of (\ref{noiseest2}) must sometimes be corrected to approach that of the noise in the signal mixture. The noise reduction itself can be done through adaptive interference cancelation (AIC), a task closely related to Acoustic Echo Cancelation (AEC), and/or postfiltering. For the latter, single-channel noise reduction methods could be used once the noise reference is given \cite{habets}.

However, all the aforementioned enhancement methods suffer from leakage of the target signal into the noise reference (\ref{bm}). This paper is therefore focused on the central problem: finding an appropriate $h$ in (\ref{noiseest1}) so that the blocking effect remains as good as possible.

\section{Survey of Known Solutions}\label{SurveySection}
\subsection{Noise-Free Conditions}\label{NoisefreeSection}
When a recording of an active target source is available in which no noise is present, the relative impulse response or the RTF can be easily estimated. Such estimates naturally provide good substitutes for $h$ in (\ref{noiseest1}).

\subsubsection{Time-domain estimation using least squares}
The mixture model (\ref{modeltime2}) without noise takes on the form 
\begin{align*}
	x_{\rm L}(n)&=s_{\rm L}(n),\\
	x_{\rm R}(n)&=\{h_{\rm rel}\ast s_{\rm L}\}(n),
\end{align*}
where $n=1,\dots,N$.
Least squares can be used to estimate the first $L$ coefficients of $h_{\rm rel}$ as
\begin{equation}\label{LSest}
{\bf h}_{\rm LS}=\arg\min_{{\bf h} \in \mathcal{R}^L} \bigl\|{\bf x}_{\rm R}-{\bf X}_{\rm L}{\bf h}\bigr\|_2^2,
\end{equation}
where ${\bf h}_{\rm LS}$ is the vector of $L$ estimated coefficients of $h_{\rm rel}$, ${\bf x}_{\rm R}=[x_{\rm R}(1-D),\dots,x_{\rm R}(N-D)]^T$ where $D$ is an integer delay due to causality, and
\[
{\bf X}_{\rm L}=
\begin{pmatrix}
x_{\rm L}(1) & 0  &\dots & 0\\
x_{\rm L}(2) & x_{\rm L}(1) & \dots & 0\\
\vdots & \ddots &\ddots & \vdots \\
x_{\rm L}(N) & x_{\rm L}(N-1) & \dots & x_{\rm L}(N-L+1) \\
0 & x_{\rm L}(N) & \dots & x_{\rm L}(N-L+2) \\
\vdots & \ddots &\ddots & \vdots \\
0 & 0 & \dots & x_{\rm L}(N) \\
\end{pmatrix}.
\]
The solution of (\ref{LSest}) is
\begin{equation}\label{LSest2}
{\bf h}_{\rm LS}=({\bf X}_{\rm L}^T{\bf X}_{\rm L})^{-1}{\bf X}_{\rm L}^T{\bf x}_{\rm R}={\bf R}^{-1}{\bf p},
\end{equation}
where
\begin{align}
{\bf R}&={\bf X}_{\rm L}^T{\bf X}_{\rm L}/N,\\
{\bf p}&={\bf X}_{\rm L}^T{\bf x}_{\rm R}/N.
\end{align}
It is worth noting that the Levinson-Durbin algorithm \cite{levinson} exploiting the Toeplitz structure of ${\bf R}$ can be used to compute ${\bf h}_{\rm LS}$ for all filter lengths $1,2,\dots,L$ in $\mathcal{O}(L^2)$ operations.
The consistency of the time-domain estimation was studied in \cite{Xu}.

\subsubsection{Frequency-domain estimation}\label{FDestimators}
The noise-free recording, in the short-term frequency-domain, takes on the form
\begin{align*}
	X_{\rm L}(\theta,\ell)&=S_{\rm L}(\theta,\ell),\\
	X_{\rm R}(\theta,\ell)&=H_{\rm RTF}(\theta)S_{\rm L}(\theta,\ell).
\end{align*}
A straightforward estimate of the RTF is given by
\begin{equation}\label{FDest}
\widehat H_{\rm RTF}(\theta)=\frac{\sum_\ell\overline{X_{\rm L}(\theta,\ell)}X_{\rm R}(\theta,\ell)}
{\sum_\ell|X_{\rm L}(\theta,\ell)|^2}.
\end{equation}

\subsection{Estimators Admitting Presence of Noise}

\subsubsection{Frequency-domain estimator using nonstationarity}
A frequency-domain estimator was proposed by Gannot et al. \cite{gannot}. It admits the presence of noise signals that are stationary or, at least, much less dynamic compared to the target signal; see also \cite{shalvi}.

The model (\ref{modelfrekv2}) can be written as
\begin{equation}\label{modelfrekv3}
\begin{split}
	X_{\rm L}(\theta,\ell)&=S_{\rm L}(\theta,\ell)+Y_{\rm L}(\theta,\ell)\\
	X_{\rm R}(\theta,\ell)&=H_{\rm RTF}(\theta) X_{\rm L}(\theta,\ell)+U(\theta,\ell)
\end{split}
\end{equation}
where $U(\theta,\ell)=Y_{\rm R}(\theta,\ell)-H_{\rm RTF}(\theta)Y_{\rm L}(\theta,\ell)$. Note that, in this form, 
$U(\theta,\ell)$ and $X_{\rm L}(\theta,\ell)$ are not independent. Let this model be valid for a certain interval during which $H_{\rm RTF}(\theta)$ is approximately constant, and let the interval be split into $P$ frames. 
By (\ref{modelfrekv3}), we have
\begin{equation}
	\Phi^p_{X_{\rm R}X_{\rm L}}(\theta)=H_{\rm RTF}(\theta)\Phi^p_{X_{\rm L}X_{\rm L}}(\theta)+
	\Phi^p_{U\,X_{\rm L}}(\theta),
\end{equation}
where $\Phi_{A\,B}^p(\theta)$ denotes the (cross) power spectral density between $A$ and $B$ during the $p$th frame.

According to the assumptions of this method (noise is stationary), $\Phi^p_{U\,X_{\rm L}}(\theta)$ should be independent of $p$ (thus written without the frame index) and the following set of equations holds
\begin{equation}\label{system2}
	\begin{bmatrix}
	\Phi^1_{X_{\rm R}X_{\rm L}}(\theta)\\
	\vdots\\
	\Phi^P_{X_{\rm R}X_{\rm L}}(\theta)
	\end{bmatrix}=
	\begin{bmatrix}
	\Phi^1_{X_{\rm L}X_{\rm L}}(\theta)&1\\
	\vdots\\
	\Phi^P_{X_{\rm L}X_{\rm L}}(\theta)&1
	\end{bmatrix}	
	\begin{bmatrix}
	H_{\rm RTF}(\theta)\\
	\Phi_{U\,X_{\rm L}}(\theta)
	\end{bmatrix}.
\end{equation}

Now, the estimate of $H_{\rm RTF}(\theta)$ is obtained by replacing the (cross-)PSDs in (\ref{system2}) by their sample-based estimates and solving the overdetermined system of equations using least squares. Theoretical analyses of bias and variance of this estimator and of the one given by (\ref{FDest}) were presented in \cite{shalvi}.

\subsubsection{Geometric Source Separation (GSS) by \cite{parra}}\label{GSSsection}
The method described here was originally designed to blindly separate directional sources whose directions of arrival (DOAs) must be given in advance (known or estimated). The method then makes use of constrained BSS so that the separating filters are kept close to a beamformer that is steering directional nulls in selected directions. We skip details of this method to save space and refer the reader to \cite{parra} or to \cite{saruwatari} for a shorter description (pages 674--675); see also a modified variant of GSS in \cite{kellermann}.

This method can be used for the RTF estimation as follows. Considering two microphones and two sources, one steered direction is selected in the DOA of the target source. The second direction is either the DOA of the (directional) interferer or, in the case of diffused or omnidirectional noise, in a direction that is apart (say 90\degree) from that of the target source. Let ${\bf W}(\theta)$ denote the resulting separating ($2\times 2$) transform that is applied to the mixed signals as
\begin{equation}\label{separated}
	{\bf y}(\theta,\ell)={\bf W}(\theta) {\bf x}(\theta,\ell),
\end{equation}
where ${\bf x}(\theta,\ell)=[X_{\rm L}(\theta,\ell) X_{{\rm R},D}(\theta,\ell)]^T$, and $X_{{\rm R},D}(\theta,\ell)$ denotes the short-term Fourier transform of $x_{\rm R}(n-D)$. Ideally, the elements of ${\bf y}(\theta,\ell)$ correspond to individual signals in the selected directions. 

Let the first row of ${\bf W}(\theta)$ be the filter that steers directional null towards the target source, which means that the first element of ${\bf y}(\theta,\ell)$ contains only noise signals. The RTF estimate is then given through
\begin{equation}\label{GSS}
	\widehat H_{\rm RTF}(\theta)=-\frac{W_{11}(\theta)}{W_{12}(\theta)},
\end{equation}
where $W_{ij}(\theta)$ denotes the $ij$th element of ${\bf W}(\theta)$.

\section{Proposed Solution}
\subsection{Motivation and Concept}
The estimators described above become biased when the assumptions used in their derivations are violated. For example, the bias in (\ref{FDest}) depends on the initial Signal-to-Noise Ratio (SNR), which may vary over time and frequency. Assuming that the SNR is sufficiently high for a given frequency, the estimator is good. But when the SNR is low, the estimator's accuracy is also low. Rather than using inaccurate estimates, we can ignore those corresponding to frequencies with low SNR values. We thus arrive at incomplete information about the RTF. That is, the estimate of $H_{\rm RTF}(\theta)$ is known only for some $\theta$s.

Based on this idea, our strategy is to construct an appropriate substitute for $h$ in (\ref{noiseest1}) using an incomplete RTF. Typical relative impulse responses are fast decaying sequences, which are compressible in the time-domain, and can thus be replaced by sparse filters \cite{lin,osher,CS,CandesNearOptimalRecovery}. These are derived through finding sparse solutions of a system built up from incomplete information in a different domain: in our case, the frequency-domain \cite{CandesDecoding,rudelson}.

We thus propose a novel method that consists of three parts\footnote{The proposed method can be modified in many ways since various solutions can be used for each part of it. We could therefore speak about a proposed class of methods. Nevertheless, the term ``proposed method'' will be used throughout the article.}:
\begin{enumerate}
	\item Pre-estimation of the RTF from a (noisy) recording. 
	\item Determination of a subset of frequencies where the estimate of the RTF is sufficiently accurate. 
	\item Computation of a sparse approximation of $h_{\rm rel}$ using the incomplete RTF.
\end{enumerate}
Various solutions can be used for each part. Potential methods to solve Part 1 have been already described in Section~\ref{SurveySection}. Part 2 can be solved in many ways depending on a given scenario, signal characteristics and the method used within Part 1; we postpone this issue to the next Section. Now we focus on a mathematical description of an appropriate method to solve Part 3.

\subsection{Nomenclature and Problem Formulation for Part 3}

Consider the Discrete Fourier Transform (DFT) domain where the length of the DFT is $M$ (sufficiently large with respect to the effective length of $h_{\rm rel}$), and, for simplicity, let $M$ be even. Let $\mathcal{S}$ denote the set of indices of frequency bins where a given RTF estimate, denoted as $\widehat H_{\rm RTF}(\theta_k)$, $k\in\mathcal{S}$ is sufficiently accurate (that is, assume that Part 1 and 2 have already been resolved). Specifically, let the values of the estimate be
\begin{equation}
	\widehat H_{\rm RTF}(\theta_k)=f_k,\quad k\in\mathcal{S}\subseteq\{1,\dots,M/2\},
\end{equation}
where $\theta_k=2k\pi/M$. 

For simplicity, the frequency bins $k=0$ and $k={M/2+1}$ can be excluded from $\mathcal{S}$ for the following symmetry to hold: Once $k\in\mathcal{S}$, then the RTF estimate is also known for $\theta_{M-k}$, namely $\widehat H_{\rm RTF}(\theta_{M-k})=\overline{f_k}$ (the conjugate value of $f_k$), since $h_{\rm rel}$ is real-valued.

Let ${\bf h}_{\rm rel}$ denote an $M\times 1$ column vector stacking $M$ coefficients of $h_{\rm rel}$, and ${\bf f}=[f_1,\dots,f_{|\mathcal{S}|}]^T$ where $|\mathcal{S}|$ denotes the cardinality of $\mathcal{S}$. The known estimates of the RTF satisfy
\begin{equation}\label{observation}
	{\bf f}={\bf F}_\mathcal{S}\cdot{\bf h}_{\rm rel}
\end{equation}
where ${\bf F}$ is the $M\times M$ matrix of the DFT, and ${\bf F}_\mathcal{S}$ is a submatrix of ${\bf F}$ comprised of rows whose indices are in $\mathcal{S}$. Since ${\bf h}_{\rm rel}$ is real, the system of linear equations (\ref{observation}) can be written as $2|\mathcal{S}|$ real-valued linear conditions 
\begin{equation}\label{observationRE}
\mathbbm{f}=\mathbbm{F}_\mathcal{S}\cdot{\bf h}_{\rm rel}
\end{equation}
where $\mathbbm{f}=[\Re({\bf f})^T\, \Im({\bf f})^T]^T$ and $\mathbbm{F}_\mathcal{S}=[\Re({\bf F}_\mathcal{S})^T\,\Im({\bf F}_\mathcal{S})^T]^T$, and $\Re(\cdot)$ and $\Im(\cdot)$ denote, respectively, the real and imaginary parts of the argument.

Since $|\mathcal{S}|$ is typically smaller than $M/2$, the system (\ref{observationRE}) is underdetermined and has many solutions. 
The key idea is to find sparse solutions that yield efficient sparse approximations of ${\bf h}_{\rm rel}$.

\subsection{Sparse solutions of (\ref{observationRE})}
The sparsest solution of (\ref{observationRE}) is defined as
\begin{equation}\label{SPA}
	{\bf g}_{\rm 0} = \arg\min_\bh 	\|{\bh} \|_0 \quad\text{w.r.t.}\quad 
		\mathbbm{f}=\mathbbm{F}_\mathcal{S}{\bf h},
\end{equation}
where $\|{\bh} \|_0$ is equal to the number of nonzero elements in $\bh$ (the $\ell_0$~pseudonorm). 
Solving this task is an NP-hard problem. Further in the paper, we will therefore consider relaxed variants based on convex programming. Several efficient greedy algorithms to solve (\ref{SPA}) exist but cannot guarantee  the finding of a global solution in general; see, e.g., \cite{tropp,cosamp}.

A more tractable formulation is based on the replacement of the $\ell_0$~pseudonorm in (\ref{SPA}) by $\ell_1$-norm, a sparsity-inducing criterion with that the optimization program becomes convex. The program is called basis pursuit \cite{bpursuit} and is defined as
\begin{equation}\label{BP}
	{\bf g}_{\rm BP} = \arg\min_\bh 	\|{\bh} \|_1 \quad\text{w.r.t.}\quad 
		\mathbbm{f}=\mathbbm{F}_\mathcal{S}{\bf h}.
\end{equation}
Using the substitution $\bh=\bh^+-\bh^-$ where $\bh^+\geq 0$ and $\bh^-\geq 0$, (\ref{BP}) can be recasted as
\begin{equation}\label{LP}
	\{{\bf g}_{\rm BP}^+, {\bf g}_{\rm BP}^- \}= \arg\min_{\bh^+,\bh^-} 	{\bf 1}^T(\bh^++\bh^-) 
\end{equation}
under the constraints
\begin{equation*}
\mathbbm{f}=\mathbbm{F}_\mathcal{S}(\bh^+-\bh^-),\quad
\bh^+\geq 0,\quad
\bh^-\geq 0,
\end{equation*}
which is indeed a linear programming problem. The solution can be found using the standard Matlab {\tt linprog} function. Other state-of-the-art optimization tools can also be used, such as the SPGL1 package\footnote{\tt http://www.cs.ubc.ca/$\sim$mpf/spgl1} by Berg et al.; see \cite{berg}.

However, neither formulation (\ref{SPA}) nor (\ref{BP}) takes into account the fact that $\mathbbm{f}$ contains certain estimation errors. It is therefore better to relax the constraint given through (\ref{observationRE}). 
One such alternative to (\ref{BP}) is LASSO (Least Absolute Shrinkage and Selector Operator) defined as
\begin{equation}
	\label{LASSO}
	{\bf g}_{\rm LASSO} = \arg\min_\bh	\|\mathbbm{F}_\mathcal{S}{\bf h}-\mathbbm{f}\|_2^2 + \tau	\|{\bh} \|_1,
\end{equation}
where $\tau\geq 0$. This formulation is closely related to the basis pursuit denoising program defined as
\begin{equation}\label{BPDN}
	{\bf g}_{\rm BPDN} = \arg\min_\bh 	\|{\bh} \|_1 \quad\text{w.r.t.}\quad 
		\|\mathbbm{F}_\mathcal{S}{\bf h}-\mathbbm{f}\|^2_2\leq\epsilon
\end{equation}
with $\epsilon\geq 0$, which is easy to interpret: The constraint $\|\mathbbm{F}_\mathcal{S}{\bf h}-\mathbbm{f}\|^2_2\leq\epsilon$ is a relaxation of $\mathbbm{f}=\mathbbm{F}_\mathcal{S}{\bf h}$ taking the possible inaccuracy in $\mathbbm{f}$ into account. LASSO is equivalent to (\ref{BPDN}) in the sense that the sets of solutions for all possible choices of $\tau$ and $\epsilon$ are the same. It means that the solution of (\ref{BPDN}) can be found through solving (\ref{LASSO}) with the corresponding $\tau$. Nevertheless, the correspondence between $\tau$ and $\epsilon$ is not trivial and is possibly discontinuous \cite{lasso2bpdn}. 

In this paper, we use a weighted formulation of (\ref{LASSO}) given by
\begin{equation}
	\label{WLASSO}
	{\bf g}_{\rm WLASSO} = \arg\min_\bh	\|\mathbbm{F}_\mathcal{S}{\bf h}-\mathbbm{f}\|_2^2 + \|{\bf w}\odot{\bh} \|_1,
\end{equation}
where ${\bf w}=[w_1,\dots,w_M]^T$ is a vector of nonnegative weights (absorbing $\tau$), and $\odot$ denotes the element-wise product. 

\begin{figure}
\centering
\includegraphics[width=0.9\linewidth]{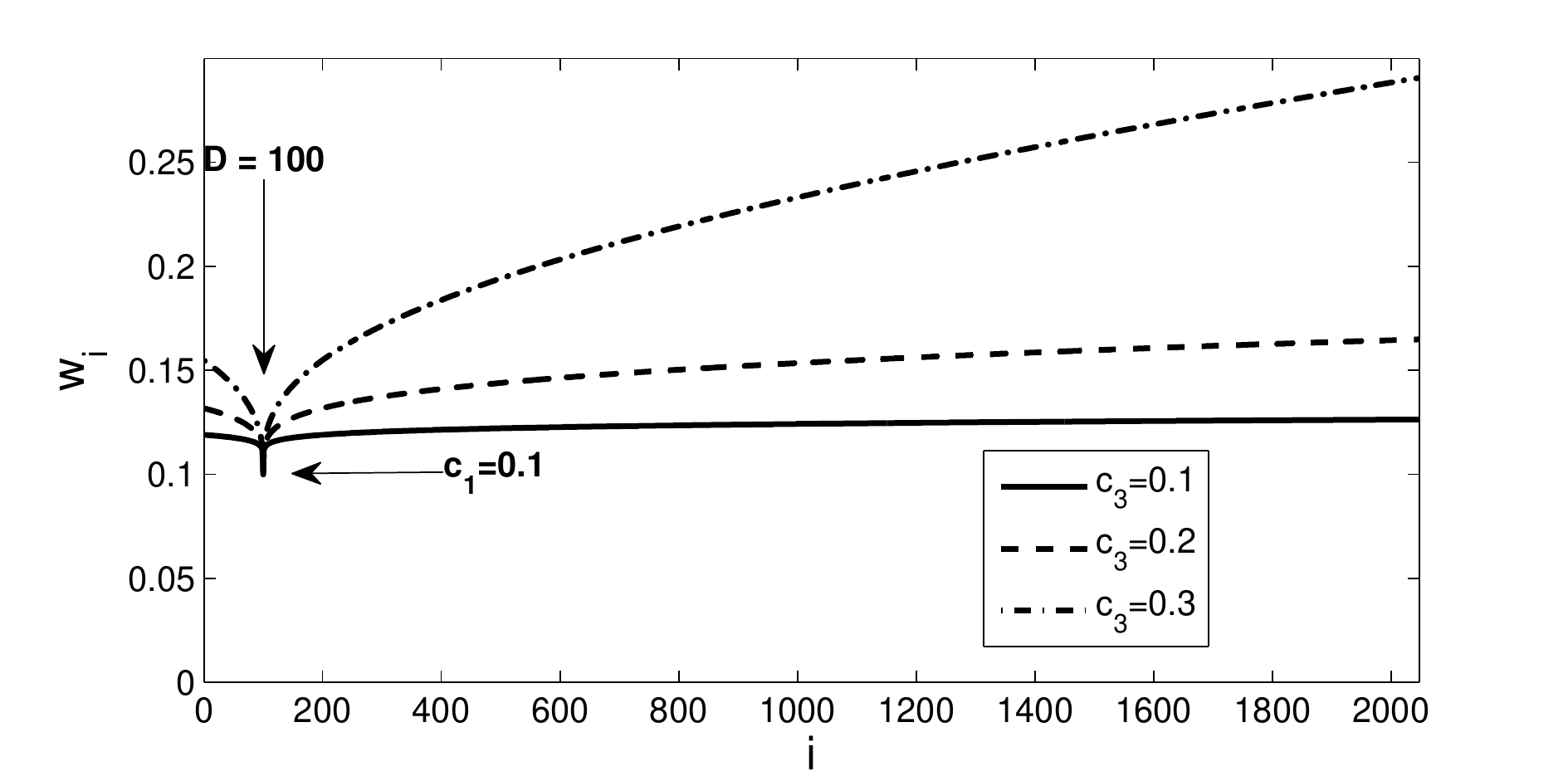}
\caption{\label{fig:figweights} Example of the weighting function (\ref{weightfun}) with $M=2048$, $D=100$, $c_1=0.1$ and $c_2=0.11$.}
\end{figure}

The weights enable us to incorporate a~priori knowledge about the solution. Elements of ${\bf g}_{\rm WLASSO}$ with higher weights tend to be closer to or equal to zero. We use this fact and select the weights to reflect the expected shape of $h_{\rm rel}$. Our heuristic choice, which is similar to that in \cite{benichoux}, is
\begin{equation}\label{weightfun}
	w_i = c_1\cdot e^{c_2|i-D|^{c_3}},\qquad i=1,\dots,M,
\end{equation}
where $c_j$, $j\in\{1,2,3\}$, are positive constants. Fig.~\ref{fig:figweights} shows three examples of this weighting function with three different values of the exponent parameter $c_3$ when $M=2048$, $D=100$, $c_1=0.1$ and $c_2=0.11$. The smallest weights are concentrated near $i=D$, because the direct-path peak of $h_{\rm rel}$ is expected there; the minimum value is $w_D=c_1$.
The weights grow with the distance from $i=D$, where the speed of the growth is controlled through $c_2$ and $c_3$. The growth of weights should reflect the expected decay in magnitudes of coefficients in $h_{\rm rel}$.

\subsection{Algorithm}
In this subsection, a proximal gradient algorithm to solve (\ref{WLASSO}) is proposed. It is a modification of SpaRSA (Sparse Reconstruction by Separable Approximation) introduced in \cite{wright}; see also closely related iterative shrinkage/thresholding methods \cite{wajs}. An advantage of these methods is their fast convergence, especially when they are initialized in the vicinity of the solution. The computational load is reduced using the properties of $\FS$.

Proximal gradient methods could be seen as a generalization of gradient descent algorithms for convex minimization programs where the objective function has the form 
\[
f({\bf h})+g({\bf h}),
\]
where both $f$ and $g$ are closed proper convex and $f$ is differentiable \cite{proximal}. Indeed, (\ref{WLASSO}) obeys this form where $f({\bf h})=\|\mathbbm{F}_\mathcal{S}{\bf h}-\mathbbm{f}\|_2^2$ and $g({\bf h})=\|{\bf w}\odot{\bh} \|_1$. 

One iteration of the proximal gradient method is
\begin{equation}\label{geniter}
	{\bf h} \leftarrow {\tt prox}_{\lambda g}\bigl({\bf h}-\lambda\nabla f({\bf h})\bigr)
\end{equation}
where 
\begin{equation}
	{\tt prox}_{\lambda g}({\bf h})=\arg\min_{\bf z}\left(g({\bf z})+ 1/(2\lambda)\|{\bf z}-{\bf h}\|_2^2\right)
\end{equation}
is the proximal operator, and $\lambda>0$ is a step-length parameter. The method is known to converge under very mild conditions; see \cite{proximal}.

By putting $f$ and $g$ from (\ref{WLASSO}) into (\ref{geniter}), we arrive at one iteration of the proposed algorithm
\begin{equation}\label{subproblem}
\bh^{t+1}=\arg\min_{\bf z} \frac{1}{2}\|{\bf z}-{\bf u}^t\|_2^2+\alpha^t\|{\bf w}\odot{\bf z}\|_1
\end{equation}
where $t=0,1,2,\dots$ is the iteration index, $\alpha^t\in[\alpha_{\rm min},\alpha_{\rm max}]$ is a variable step-length parameter, and 
\begin{equation}\label{ut}
{\bf u}^t=\bh^t-\alpha^t\FS^T(\FS\bh^t-\fff).
\end{equation}
The elements of ${\bf z}$ are separable in (\ref{subproblem}), which allows us to find the solution in closed-form \cite{wright}, that is 
\begin{equation}\label{sparsa}
	{\bh}^{t+1}={\tt soft}({\bf u}^t,\alpha^t{\bf w})
\end{equation}
where ${\tt soft}(u,a)={\tt sign}(u)\max\{|u|-a,0\}$. In (\ref{sparsa}), this ``soft-thresholding'' function is applied element-wise. 

The step-length parameter $\alpha^t$ is chosen as in SpaRSA
\begin{equation}\label{stepsparsa}
\alpha^t=\frac{\|{\bf h}^t-{\bf h}^{t-1}\|_2^2}{\|\FS({\bf h}^t-{\bf h}^{t-1})\|_2^2},
\end{equation}
which was derived based on a variant of the Barzilai-Borwein spectral approach; see \cite{wright}.

To terminate the algorithm, we derive a stopping criterion as follows. It holds that ${\bf g}_{\rm WLASSO}$ is the solution of (\ref{WLASSO}) if and only if it satisfies \cite{romberg}
\begin{align}
(\FS)_\Gamma^T(\FS\cdot{\bf g}_{\rm WLASSO}-\fff)&=-{\bf w}_\Gamma\odot {\bf q}_\Gamma,\label{cond1}\\
\bigl|(\FS)_{\Gamma^c}^T(\FS\cdot{\bf g}_{\rm WLASSO}-\fff)\bigr|&<{\bf w}_{\Gamma^c},\label{cond2}
\end{align}
where the subscript $(\cdot)_\Gamma$ denotes the restriction to indices (columns in the case of a matrix) in the set $\Gamma$; ${\bf q}$  is the vector of signs of ${\bf g}_{\rm WLASSO}$, that is ${\bf q}={\tt sign}({\bf g}_{\rm WLASSO})$; $\Gamma$ is the set of indices of nonzero elements of ${\bf g}_{\rm WLASSO}$ (the active set), and $\Gamma^c$ is its complement to $\{1,\dots,M\}$. We define the termination criterion that assesses the degree of validity of (\ref{cond1}) as
\begin{equation}
	{\tt crit}(\bh^t)=\Bigl\|\bigl(\FS^T(\FS\cdot\bh^t-\fff)+{\bf w}\odot{\tt sign}(\bh^t)\bigr)_\Gamma\Bigr\|_2^2.
\end{equation}
The algorithm stops iterating when ${\tt crit}(\bh^t)\leq{\tt tol}$ where ${\tt tol}$ is a small positive constant.

Using the fact that ${\bf g}_{\rm WLASSO}$ satisfies (\ref{cond1}) and (\ref{cond2}), it can be shown that ${\bf g}_{\rm WLASSO}$ is a fixed point of (\ref{subproblem}). The global convergence of the algorithm (although with a different stopping criterion) was proven in \cite{wright}.

Most of the computational burden is due to the vector-matrix products by $\FS$ and $\FS^T$ in (\ref{ut}) and in (\ref{stepsparsa}). Since $\FS$ only represents a part of the DFT, the products can be computed via the (inverse) Fast Fourier transform, which also leads to memory savings as $\FS$ is determined only through $\mathcal{S}$. The computational complexity of one iteration is thus $\mathcal{O}(M\log M)$. A pseudo-code of the algorithm\footnote{The Matlab implementation of Algorithm~1 is available at {\tt http://itakura.ite.tul.cz/zbynek/downloads.htm}} is summarized in Algorithm~\ref{spaFFT}. 

\begin{algorithm}[t]\label{spaFFT}
{\footnotesize
 \caption{Algorithm to solve (\ref{WLASSO})}
 \SetAlgoLined
 \KwIn{$\mathcal{S}$, $\mathbbm{f}$, ${\bf w}=[w_1,\dots,w_M]^T$, ${\bf h}^0$}
 \KwOut{$\bh^t$}
${\bf d}={\bf 0}_{M\times 1}$, ${\bf r}^0=\FS{\bf h}^0-\mathbbm{f}$, $\nabla^{0}=\FS^T{\bf r}^0$, $i=\sqrt{-1}$\\
$t=0$\\
\While{${\tt crit}(\bh^{t})>{\tt tol}$}
{
${\bf h}^{t+1} = {\tt soft}({\bf h}^{t} - \alpha\nabla^{t},\alpha{\bf w})$\\
$\Delta{\bf h} = {\bf h}^{t+1}-{\bf h}^{t}$\\
${\bf a} = {\tt fft}(\Delta{\bf h})$\\
${\bf b} = \bigl[\Re({\bf a}_\mathcal{S})^T\,\Im({\bf a}_\mathcal{S})^T\bigr]^T$ \tcc*[f]{now ${\bf b}=\FS\Delta{\bf h}$}\\
${\bf r}^{t+1} = {\bf r}^{t} + {\bf b}$ \tcc*[f]{now ${\bf r}^{t+1}=\FS{\bf h}^{t+1}-\mathbbm{f}$}\\
$\alpha^{t+1} = \min(\alpha_{\rm max},\max(\alpha_{\rm min},\|\Delta{\bf h}\|_2^2/\|{\bf b}\|_2^2))$\\
${\bf d}_\mathcal{S}={\bf r}^{t+1}_{1:|\mathcal{S}|} + i {\bf r}^{t+1}_{|\mathcal{S}|+1:2|\mathcal{S}|}$\\
$\nabla^{t+1}=M/2\cdot{\tt ifft}({\bf d},M,\text{\tt 'symmetric'})$\\ \tcc*[f]{now $\nabla^{t+1}=\FS^T{\bf r}^{t+1}$}\\ 
$t\leftarrow t+1$\\
}
 }
\end{algorithm}

\section{Determining the Set $\mathcal{S}$}\label{setSsection}

This Section is dedicated to solutions of Part 2 of the proposed method.
Let the estimates $\widehat H_{\rm RTF}(\theta_k)$ of $H_{\rm RTF}(\theta_k)$ be given for all $k$. The task is to select the set $\mathcal{S}$ such that $\widehat H_{\rm RTF}(\theta_k)$ is sufficiently accurate for $k\in\mathcal{S}$. 

\subsection{Oracle Inference}\label{oracleinference}
For experimental purposes, we define an oracle method that comes from complete knowledge of the SNR in the frequency domain. For simplicity, we can consider the SNR on the left microphone only, which is given by
\[
\mbox{SNR}_{\rm L}(\theta_k)=\frac{\sum_\ell|S_{\rm L}(\theta_k,\ell)|^2}{\sum_\ell|Y_{\rm L}(\theta_k,\ell)|^2}.
\]
This method selects frequencies for which the SNR is higher than a positive adjustable parameter $\beta$. The resulting set will be denoted as $\mathcal{S}^{\rm or}_\beta$. Specifically, it holds that
\begin{equation}\label{oraclechoice}
	k\in\mathcal{S}^{\rm or}_\beta \, \Longleftrightarrow \, \mbox{SNR}_{\rm L}(\theta_k)>\beta. 
\end{equation}

Now we focus on methods that do not require prior knowledge of SNR.

\subsection{Kurtosis-Based Selection}
For cases where the target signal is a speaker's voice while the other sources are non-speech, voice activity detectors (VAD) can be used to infer high-SNR frequency bins \cite{tashev}. Here we use a simple detector based on kurtosis. Kurtosis is often used as a contrast function reflecting (non)-Gaussian character of a random variable, because the kurtosis of a Gaussian variable is equal to zero. 
For example, a VAD using kurtosis was proposed in \cite{nemer}; a recent method for blind source extraction using kurtosis was proposed in \cite{cichocki}.

For a complex-valued random variable $X$, normalized kurtosis is defined as
\begin{equation}
	{\tt kurt}(X)=\frac{\mbox{E}[|X|^4]- |\mbox{E}[X^2]|^2}{\mbox{E}[|X|^2]^2}-2,
\end{equation}
where $\mbox{E}[\cdot]$ stands for the expectation operator, which is replaced by the sample mean in practice.
Speech signals often yield positive kurtosis. We therefore define the set of selected frequencies as
\begin{equation}\label{kurtosischoice}
	k\in\mathcal{S}^{\rm kurt}_\beta\, \Longleftrightarrow\, {\tt kurt}\bigl(X_{\rm L}(\theta_k,\ell)\bigr)>\beta.
\end{equation}
In other words, frequencies that yield higher kurtosis than $\alpha$ on the left channel are supposed to contain a dominating target (speech) signal. 

\subsection{Selection Methods after applying BSS}

\subsubsection{Divergence}
Some BSS methods, such as GSS described in Section~\ref{GSSsection}, proceed by numerical optimization of a contrast function that evaluates the independence of separated outputs. For example, GSS minimizes a criterion for approximate joint diagonalization of covariance matrices of the input signals computed on frames, plus a penalty function ensuring a constraint \cite{parra}. When the minimum of the function is shallow, the convergence is slow, which might be indicative of 
poor separation. 

Therefore, the method proposed here rejects frequencies for which the algorithm did not converge within a selected number of iterations. Thus, the selection is
\begin{equation}\label{divergencechoice}
	k\in\mathcal{S}^{\rm div}_Q\, \Longleftrightarrow\, \text{${\bf W}(\theta_k)$ converged within $Q$ iterations}.
\end{equation}

\subsubsection{Coherence-Based Selection}
Another way to assess the separation quality without knowing the achieved SNR is to compute the coherence function among the separated signals. As the separated signals should be independent, the coherence, defined as
\begin{equation}
	{\tt coh}(\theta_k)=\frac{|\sum_\ell y_1(\theta_k,\ell) \overline{y_2(\theta_k,\ell)}|}
	{\sqrt{\sum_\ell |y_1(\theta_k,\ell)|^2}\sqrt{\sum_\ell |y_2(\theta_k,\ell)|^2}}
\end{equation}
should be ``small''. Here, $y_i(\theta_k,\ell)$ denotes the $i$th separated signal, that is, the $i$th element of ${\bf y}(\theta_k,\ell)$ defined in (\ref{separated}). Now, the selection is defined as
\begin{equation}\label{coherencechoice}
	k\in\mathcal{S}^{\rm coh}_\beta\, \Longleftrightarrow\, {\tt coh}(\theta_k)<\beta.
\end{equation}

\subsection{Thresholds}
Note that there is no clear correspondence between the values of $\beta$s in (\ref{oraclechoice}), (\ref{kurtosischoice}) and (\ref{coherencechoice}). Rather than determining values for these parameters, $\beta$s will be chosen based on a pre-selected ratio of accepted frequencies in percents (this quantity will later be referred to as {\em percentage}).

\section{Experiments}
\begin{figure}
\centering
\includegraphics[width=0.8\linewidth]{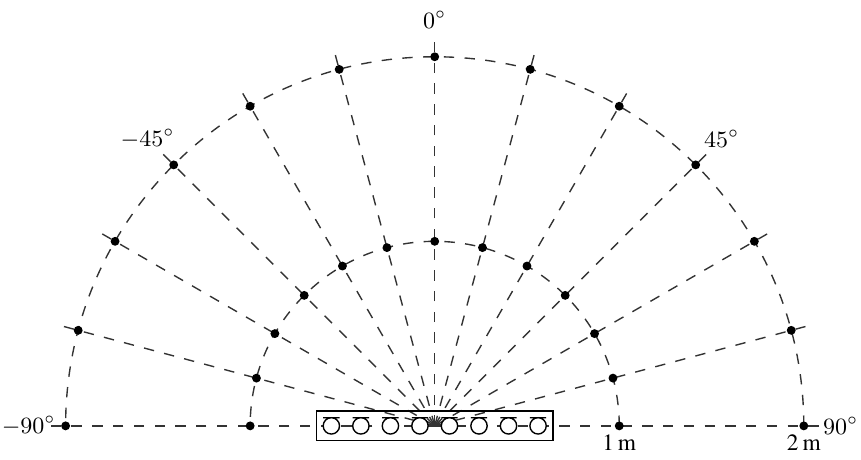}
\caption{\label{BAR} Illustration of the geometric setup of impulse response database from \cite{irdatab}. The picture is a reprint from \cite{irdatab} with the permission of its authors.}
\end{figure}

We present results of experiments evaluating and comparing the ability of several methods to attenuate a target speaker in noisy stereo recordings. Each scenario is simulated using a database\footnote{\tt http://www.eng.biu.ac.il/gannot/downloads/} of room impulse responses (RIR) measured in the speech \& acoustic lab of the Faculty of Engineering at Bar-Ilan University \cite{irdatab}. The lab is a $6\times 6 \times 2.4$~m room with variable
reverberation time (T$_{60}$ is set, respectively, to 160~ms, 360~ms and 640~ms). The database consists of impulse responses relating eight microphones and a loudspeaker. The microphones are arranged to form a linear array (we use pairs of microphones from the arrangement $3-3-3-8-3-3-3$~cm) and the loudspeaker is placed  at various angles from $-90$ to $90\degree$ at distances of $1$ and $2$~m; see the setup depicted in Fig.~\ref{BAR}. All computations were done in Matlab$^{TM}$ on a standard PC with four-core processor 2.6~GHz and 8~MB of RAM.

Noise signals are either diffused and isotropic (shortly referred to as omnidirectinal) or simulated to be directional (one channel of an original noise signal is convolved with RIRs corresponding to the interferer's position). Sample of omnidirectional babble noise is taken from the database recorded in the lab. 
Signals for directional sources are taken from the task of the SiSEC~2013 evaluation campaign \cite{sisec2013}\footnote{\tt http://sisec.wiki.irisa.fr/} titled ``Two-channel mixtures of speech and real-world background noise.'' We use a female and a male utterance and a sample of babble noise recorded in a cafeteria\footnote{This sample is used to simulate a directional babble noise although typical babble noise is diffused and isotropic. The purpose of this sample is to also have another directional source besides the Gaussian noise.}. The signals are $10$~s long, and the sampling frequency is $16$~kHz.

Once microphone responses of the sources are prepared, they are mixed together at a specified SNR averaged over both microphones. Specifically,
\begin{equation}\label{iniSNR}
	\mbox{SNR}_{\rm in}=\frac{\sum_{i\in\{{\rm L,R}\}}\sum_n [\{h_i\ast s\}(n)]^2}{\sum_{i\in\{{\rm L,R}\}}\sum_n [y_i(n)]^2},
\end{equation}
where $n$ spans a given interval of data. 

The testing sample (10~s) is split into intervals with 75\% overlap; experiments are always conducted on each interval (37 independent trials when the interval length is 1~s) and the results are averaged. For a particular interval, SNR at the output of (\ref{bm}) is computed as
\begin{equation}\label{outSNR}
	\mbox{SNR}_{\rm out}=\frac{\sum_n[\{g\ast s_{\rm L}\}(n)- s_{\rm R}(n)]^2}{\sum_n[\{g\ast y_{\rm L}\}(n)-y_{\rm R}(n)]^2},
\end{equation}
where $s_{\rm R}=h_{\rm R}\ast s$ (the response of the target signal on the right microphone), and $g$ denotes the estimate of $h_{\rm rel}$. The numerator of (\ref{outSNR}) corresponds to the leakage of the target signal in (\ref{bm}) while the denominator contains the desired noise reference.

The final criterion is the {\em attenuation rate}  evaluated as the ratio between $\mbox{SNR}_{\rm out}$ and $\mbox{SNR}_{\rm in}$. The more negative the value (in dBs) of this criterion is, the better the evaluated filter performs.

We compare several variants of the proposed method combining different approaches to solve Part 1 and Part 2; Part 3 is the same in all instances. The methods used in Part 1 (FD, NSFD and GSS) are always compared with those obtained after Parts 2 and 3, as the main goal is that the latter improve the former; see the list of compared methods in Table~\ref{table:variants}.


\begin{table*}
\centering
\caption{\label{table:variants} Methods Compared in Experiments}
\begin{tabular}{l|l|l}
	 Method Acronym & Method used in Part~1 & Method used in Part~2 \\
	\hline	
	FD & frekv.-dom. estimator (\ref{FDest}) & - \\
	FD$^\text{or}$ & frekv.-dom. estimator (\ref{FDest}) & oracle selector (\ref{oraclechoice})\\
	FD$^\text{kurt}$ & frekv.-dom. estimator (\ref{FDest}) & kurtosis-based selector (\ref{kurtosischoice})\\
	NSFD & non-stationarity-based frekv.-dom. estimator (\ref{system2}) & -  \\
	NSFD$^\text{or}$ & non-stationarity-based frekv.-dom. estimator (\ref{system2}) & oracle selector (\ref{oraclechoice}) \\
	NSFD$^\text{kurt}$ & non-stationarity-based frekv.-dom. estimator (\ref{system2}) & kurtosis-based selector (\ref{kurtosischoice}) \\
	GSS & BSS estimator (\ref{GSS}) & - \\
	GSS$^\text{div}$ & BSS estimator (\ref{GSS}) & divergence-based selector (\ref{divergencechoice}) \\
	GSS$^\text{coh}$ & BSS estimator (\ref{GSS}) & coherence-based selector (\ref{coherencechoice})\\
\end{tabular}
\end{table*}

If not specified otherwise, parameters are set to the default values shown in Table~\ref{table:defaultexp}. Note that microphone distances are differently selected for FD and NSFD and for GSS in order to provide setups that are preferable for each method (optimized based on the results).

\begin{table}
\centering
\caption{\label{table:defaultexp} Default Settings in Experiments}
\begin{tabular}{l|c}
	 Parameter & Value [units]\\
	\hline	
	Sampling frequency & 16~kHz\\
	Data interval length per trial & 1~s\\
	T$_{60}$ & 360 ms \\
	SNR$_{\rm in}$ & 0 dB \\
	Target angle & $0\degree$ \\
	Directional interferer angle & $-60\degree$ \\
	Distance of sources to microphones & 2~m \\
	Length of DFT $M$ & 2048 \\
	Window shift in short-term DFT & 64 \\
	Delay parameter $D$ & 100 \\
	Microphone pair when using FD or NSFD & [3 4] (3 cm)\\
	Microphone pair when using GSS & [4 5] (8 cm)\\
	$c_1$, $c_2$, $c_3$ in (\ref{weightfun}) & 0.1, 0.11, 0.3	\\
	Frame length in NSFD & 1000\\
	Number of blocks in GSS & 4\\
	$\alpha_{\rm min}$, $\alpha_{\rm max}$, ${\tt tol}$ & $10^{-7}$, $10^3$, $10^{-3}$\\
\end{tabular}
\end{table}

\subsection{Attenuation Rate vs. Percentage}\label{experiment1}
The number of selected frequencies within Part 2 (the parameter we refer to as the {\em percentage}) has a particular influence on the resulting estimator\footnote{Results of methods that do not allow the choice of the percentage are in graphs shown as constant lines.}. 
On the one hand, the attenuation rate is always poor when the percentage is lower than a certain threshold (depending on the method and experiment). On the other hand, the rate is always getting back to that of the initial estimator as the percentage approaches 100\%. It is desirable  that the rate should be improved, at least for some values in between these two extremes.

\subsubsection{Diffused and isotropic noise}
\begin{figure}
	\centering
		\includegraphics[width=0.5\linewidth]{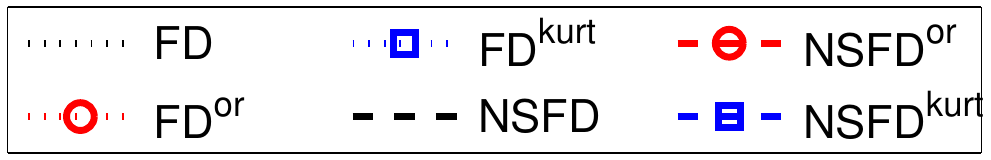}\\
		\includegraphics[width=0.49\linewidth]{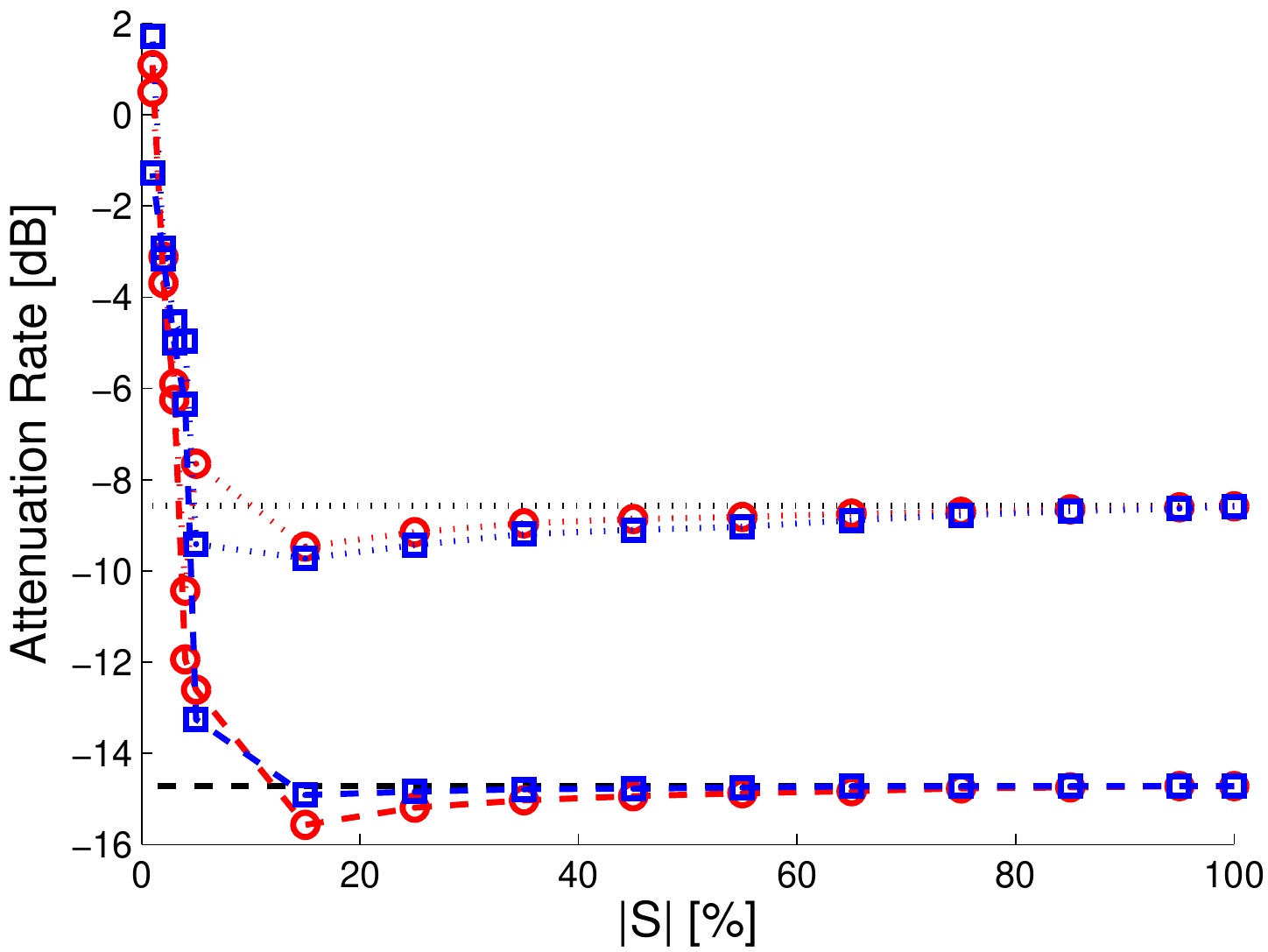}
		\includegraphics[width=0.49\linewidth]{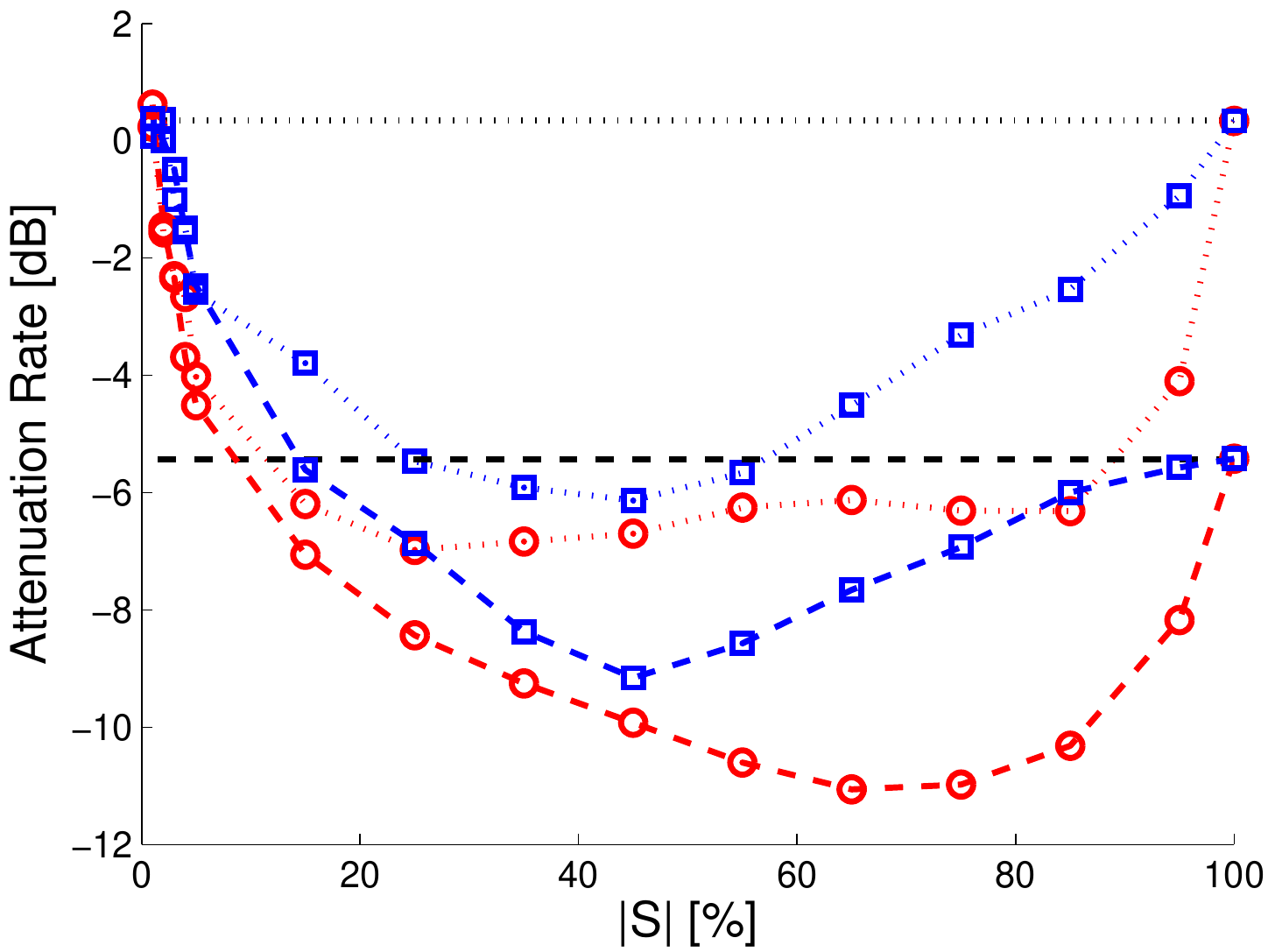}\\
		(a)\hspace{0.5\linewidth}(b)
		\caption{\label{fig:noise_diffus} Female target speaker interfered by (a) Gaussian stationary and spatially and temporally white noise and (b) omnidirectional babble noise.}
\end{figure}

\begin{figure*}
	\centering
		\includegraphics[width=0.99\linewidth]{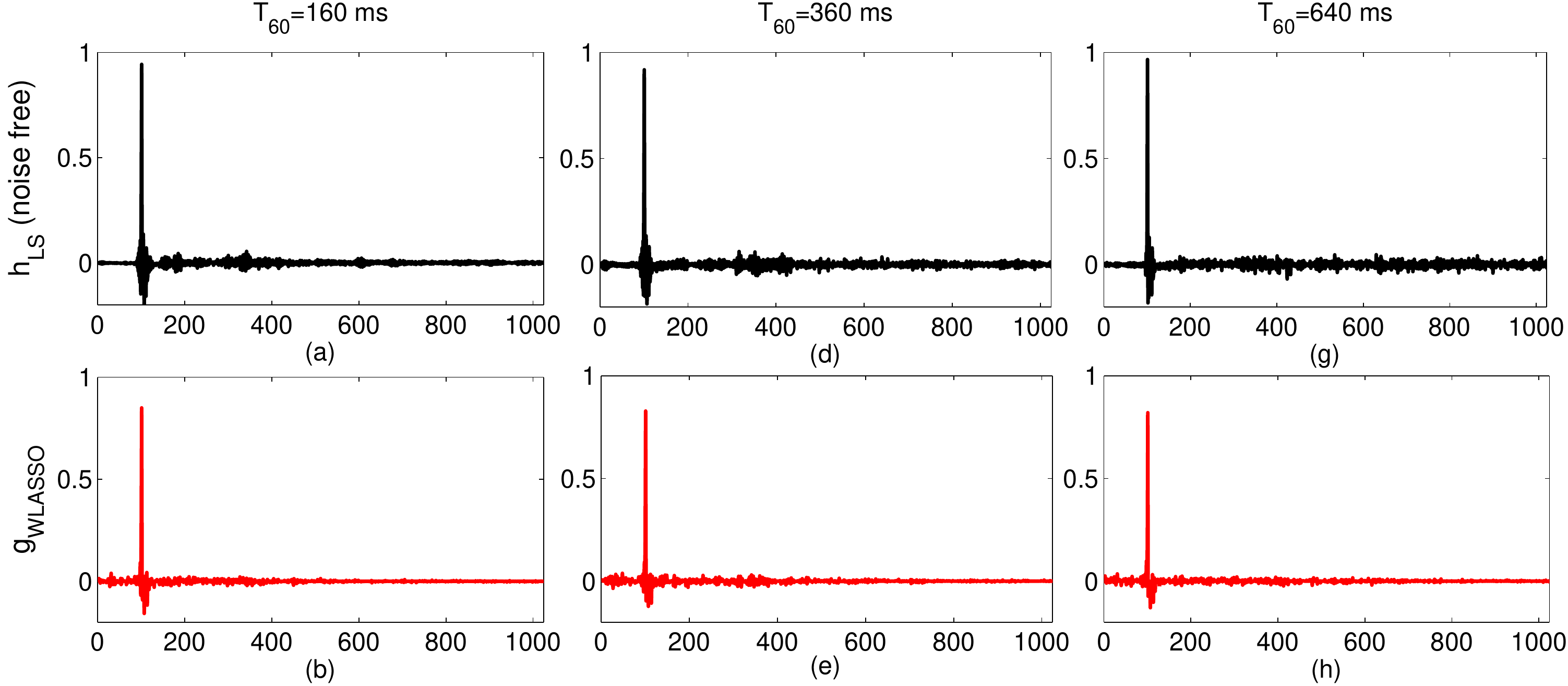}\\
		\includegraphics[width=0.99\linewidth]{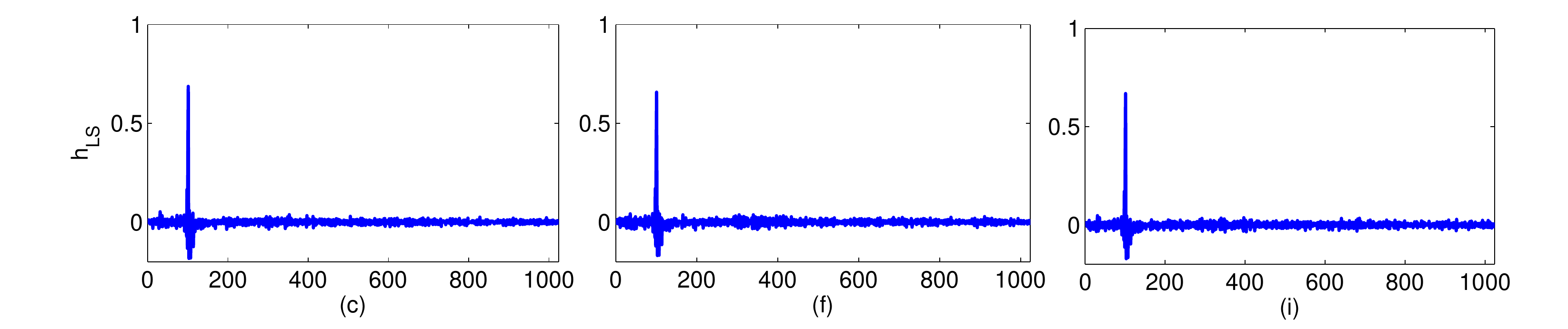}
		\caption{\label{fig:hrelexam} Examples of ReIRs computed in the first trial of the experiment of Section~\ref{experiment1} for three different reverberation times (columns) when the female target speaker was interfered by omnidirectional babble noise. The first row contains the least-squares estimates according to (\ref{LSest2}) from noise-free recording of the target while the third row contains the estimates computed from noisy data. The second row contains the sparse approximations computed from $50$\% incomplete RTF estimate by NSFD$^{\rm kurt}$ (from noisy data). The attenuation rates by the estimated ReIRs were, respectively, (a) -22.4~dB, (b) -12.8~dB, (c) -8.6, (d) -23.7~dB, (e) -11.0~dB, (f) -8.0~dB, (g) -14.7~dB, (h) -7.1~dB, and (i) -6.5~dB.}
\end{figure*}

Figures \ref{fig:noise_diffus}(a) and \ref{fig:noise_diffus}(b) show results from two experiments when the target signal (female speech) is contaminated, respectively, by stationary Gaussian white noise that is spatially white (independently generated on each channel) and by the omnidirectional babble noise. 

The white noise situation (Fig.~\ref{fig:noise_diffus}(a)) favors NSFD as it obeys the assumed model \cite{gannot}. Now NSFD$^\text{or}$ and NSFD$^\text{kurt}$ perform approximately the same as NSFD or marginally improve the attenuation rate (maximum by 1~dB) unless the percentage goes below 15\%.  
The methods based on FD behave similarly but do not outperform those based on NSFD. The original NSFD is hard to outperform in this scenario as its performance is close to optimal. 

In babble noise, NSFD attenuates the target by about $5$~dB, while FD yields an attenuation rate above $0$~dB, and hence fails. The proposed methods successfully improve these results for a wide range of the percentage values. The best improvements are achieved through oracle methods NSFD$^\text{or}$ (70\%) and FD$^\text{or}$ (20--80\%), where the attenuation rates by NSFD and FD are improved by about $6$~dB. The optimum improvement by the kurtosis-based variants NSFD$^\text{kurt}$ (45\%) and FD$^\text{kurt}$ (45\%) is by 4-6~dB, which is only reasonably lower compared to that of the oracle-based frequency selections.
The results confirm that the kurtosis-based selection is efficient in detecting frequencies with high SNR when the noise is Gaussian or babble. Examples of estimated ReIRs in this experiment are shown in Fig.~\ref{fig:hrelexam}.

We also examined the case when the target source was shifted to a 60\degree angle. The results, not shown here due to space constraints, were comparable with the results for 0\degree.

\subsubsection{Directional noise}
\begin{figure}
	\centering
	\includegraphics[width=0.5\linewidth]{legend_noGSS.pdf}\\
		\includegraphics[width=0.49\linewidth]{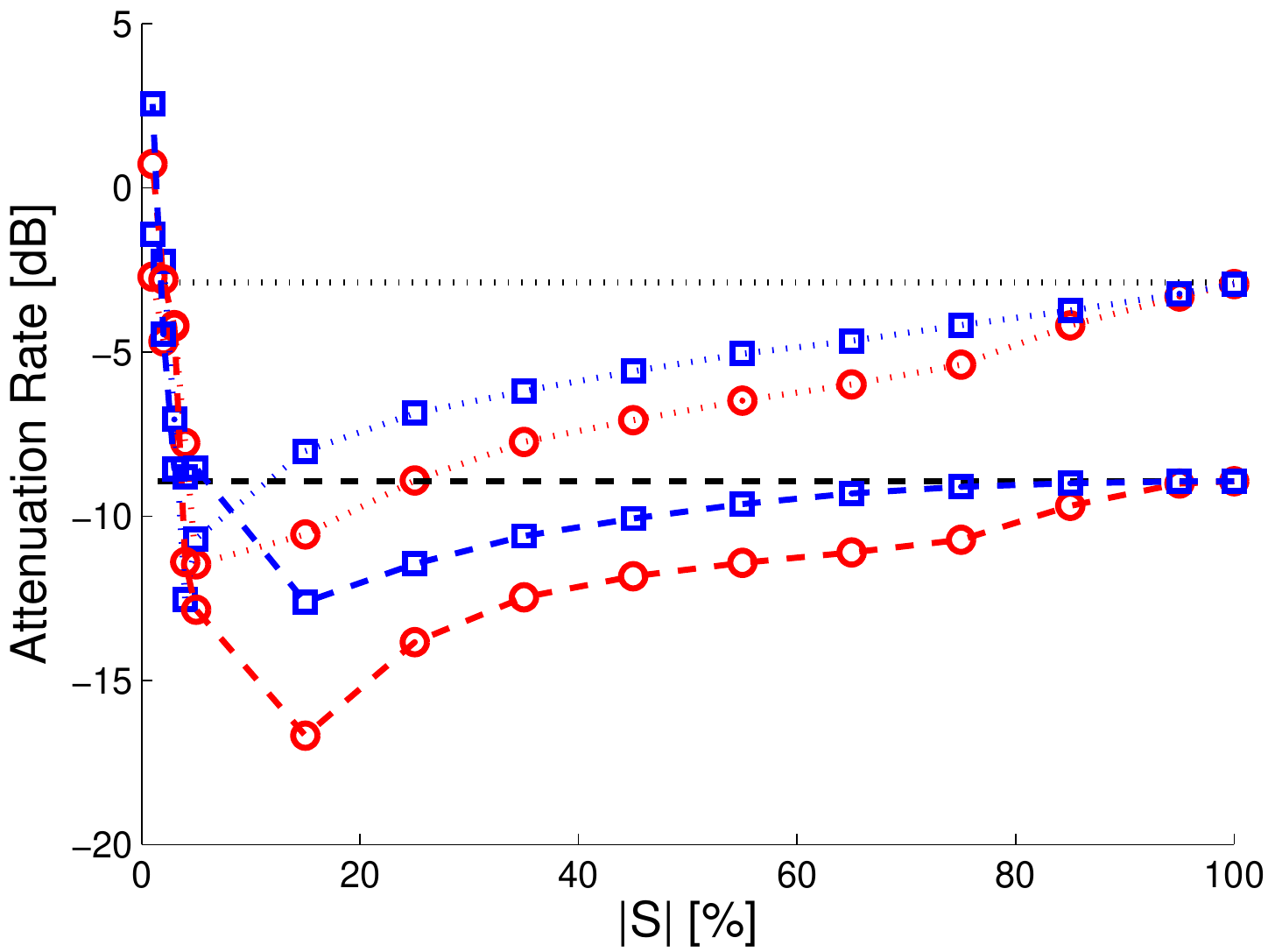}
		\includegraphics[width=0.49\linewidth]{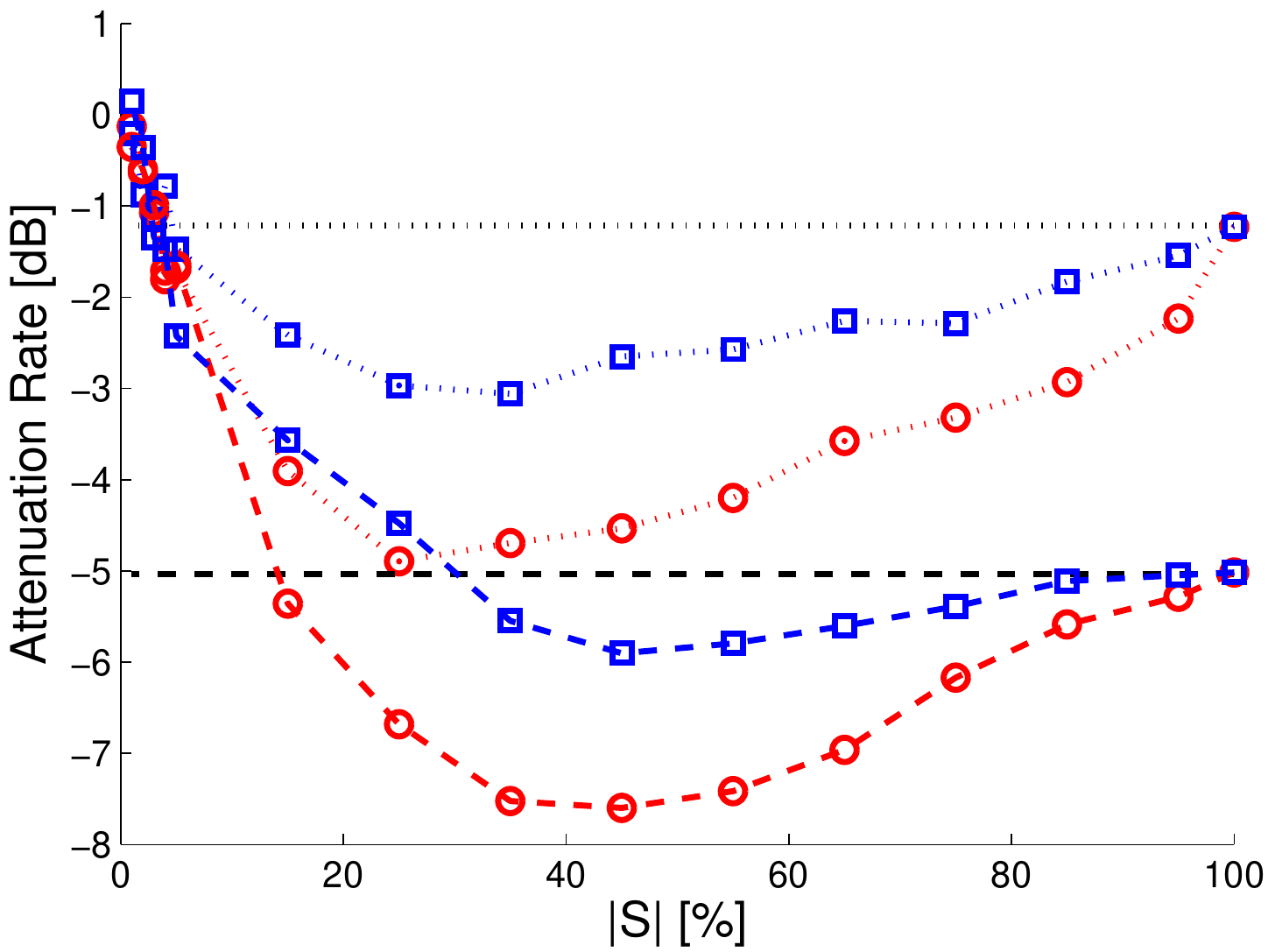}\\
		(a)\hspace{0.5\linewidth}(b)\\
		\includegraphics[width=0.49\linewidth]{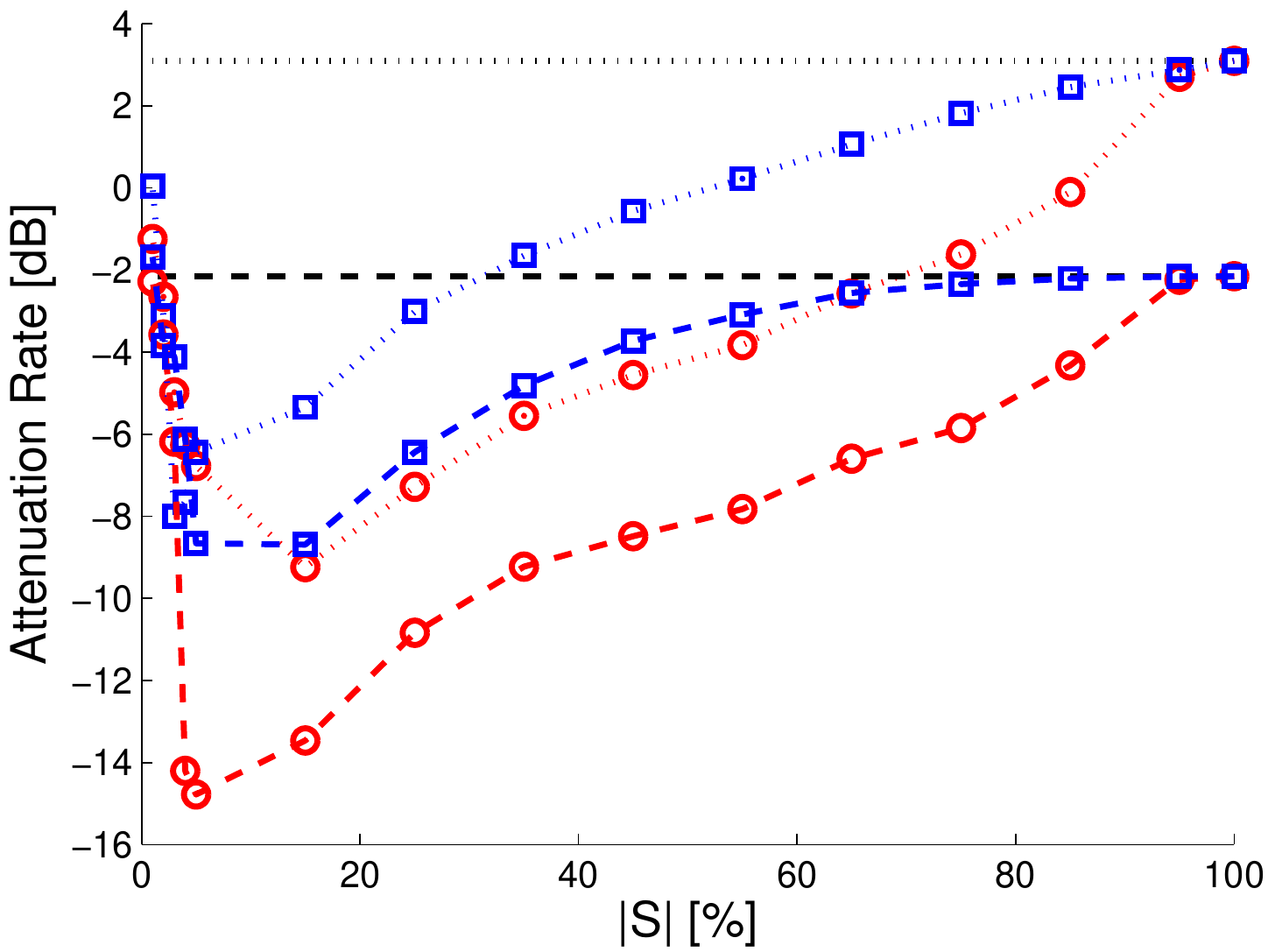}
		\includegraphics[width=0.49\linewidth]{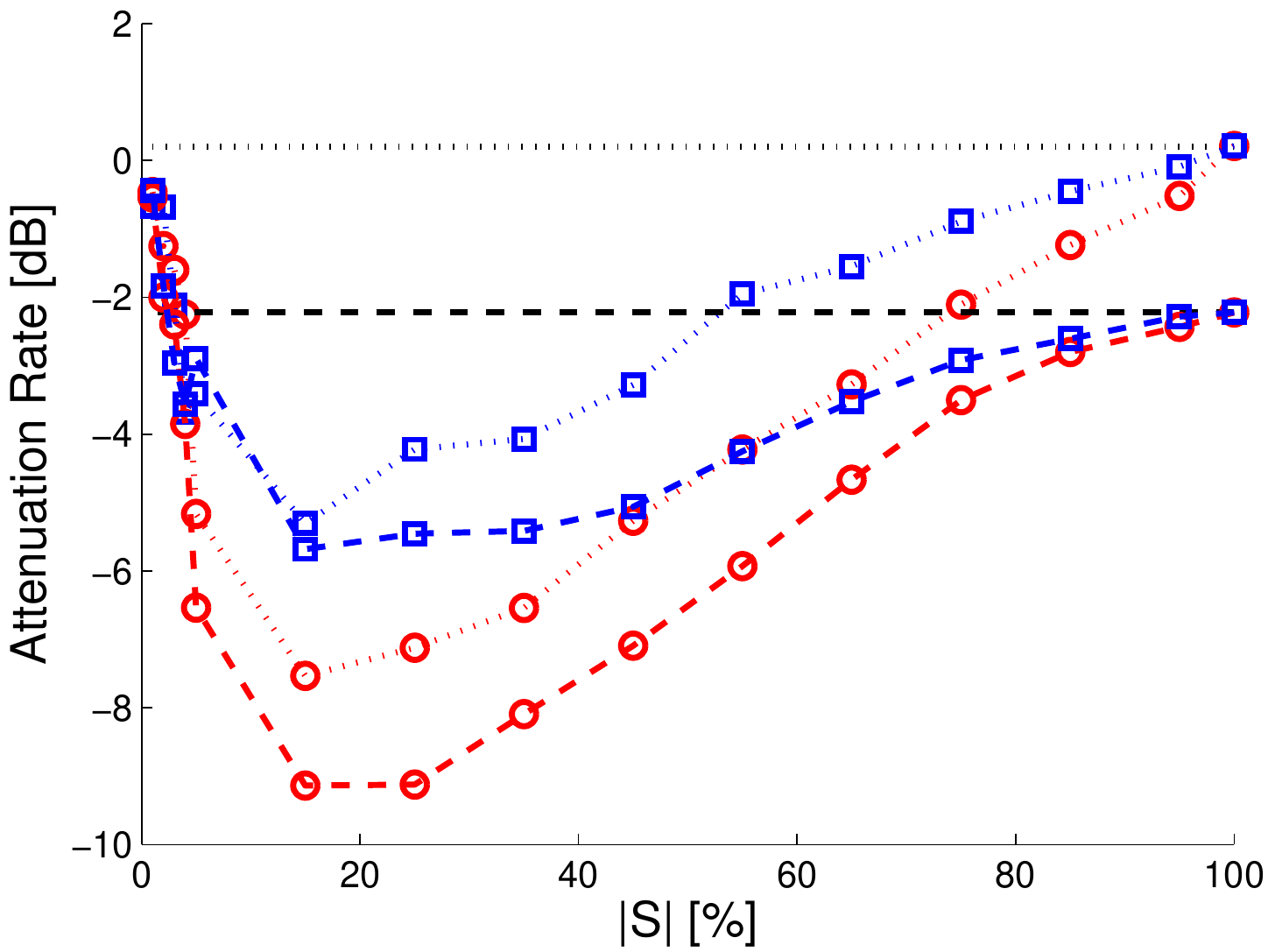}\\
		(c)\hspace{0.5\linewidth}(d)
		\caption{\label{fig:noise_direct} Results of the experiment where the target source at angle $\beta$ is interfered by directional noise from $-60\degree$: (a) Gaussian noise and $\beta=0\degree$, (b) babble noise and $\beta=0\degree$, 
		(c) Gaussian noise and $\beta=60\degree$, (d) babble noise and $\beta=60\degree$.}
\end{figure}

Fig.~\ref{fig:noise_direct} shows results of experiments when noise signals were played from a loudspeaker placed at $-60\degree$ and the target was placed at an angle of $0\degree$ or $60\degree$. 

By comparing Fig.~\ref{fig:noise_diffus}(a) with Figures~\ref{fig:noise_direct}(a) and \ref{fig:noise_direct}(c), FD and NSFD perform worse by 5--6~dB and by 11--13~dB, respectively, when the Gaussian noise is directional and the target speaker stands at angles of $0\degree$ and $60\degree$. This means the directional noise scenario is now less favorable for both FD and NSFD than in the previous scenario.  To explain, note that within the frequency bins with low activity of the target source, these methods, in fact, estimate the RTF of the (directional) noise source. When applying such estimated RTF to attenuate the target signal, part of the noise source is attenuated as well, which causes loss in terms of the attenuation rate. 

It should also be noted that the performance loss may be even higher when the target is spatially more separated from the noise source ($60\degree$), because the higher the spatial separation of the directional noise source, the higher the bias in the RTF estimates by FD and NSFD could be.

NSFD$^\text{or}$ and NSFD$^\text{kurt}$ as well as FD$^\text{or}$ and FD$^\text{kurt}$ improve their initial methods, especially when the percentage value approaches 15\%. Moreover, these methods yield an attenuation rate that is close to that achieved with the spatially white Gaussian noise in Fig.~\ref{fig:noise_diffus}(a). Compared to FD and NSFD, the proposed methods do not attenuate the directional noise in the frequency bins with low target source activity.

Similar, but not identical, conclusions can be drawn for the babble noise case. The results by NSFD in Fig.~\ref{fig:noise_direct}(b) are almost the same as those in Fig.~\ref{fig:noise_diffus}(b), while, in Fig.~\ref{fig:noise_direct}(d), the attenuation by NSFD drops by 3~dB compared to Fig.~\ref{fig:noise_diffus}(b).

\subsubsection{A speaking interferer}\label{speakersexperiment}

\begin{figure}
	\centering
	\includegraphics[width=0.5\linewidth]{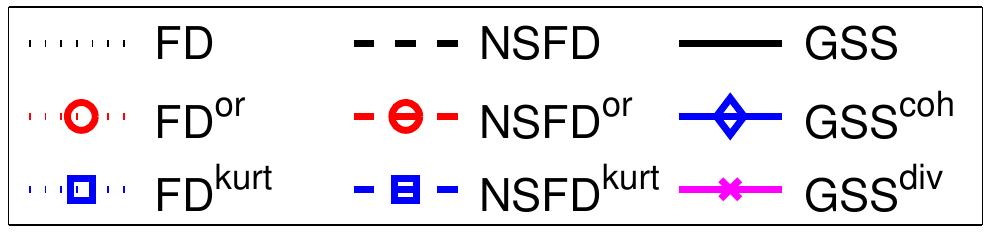}\\
		\includegraphics[width=0.9\linewidth]{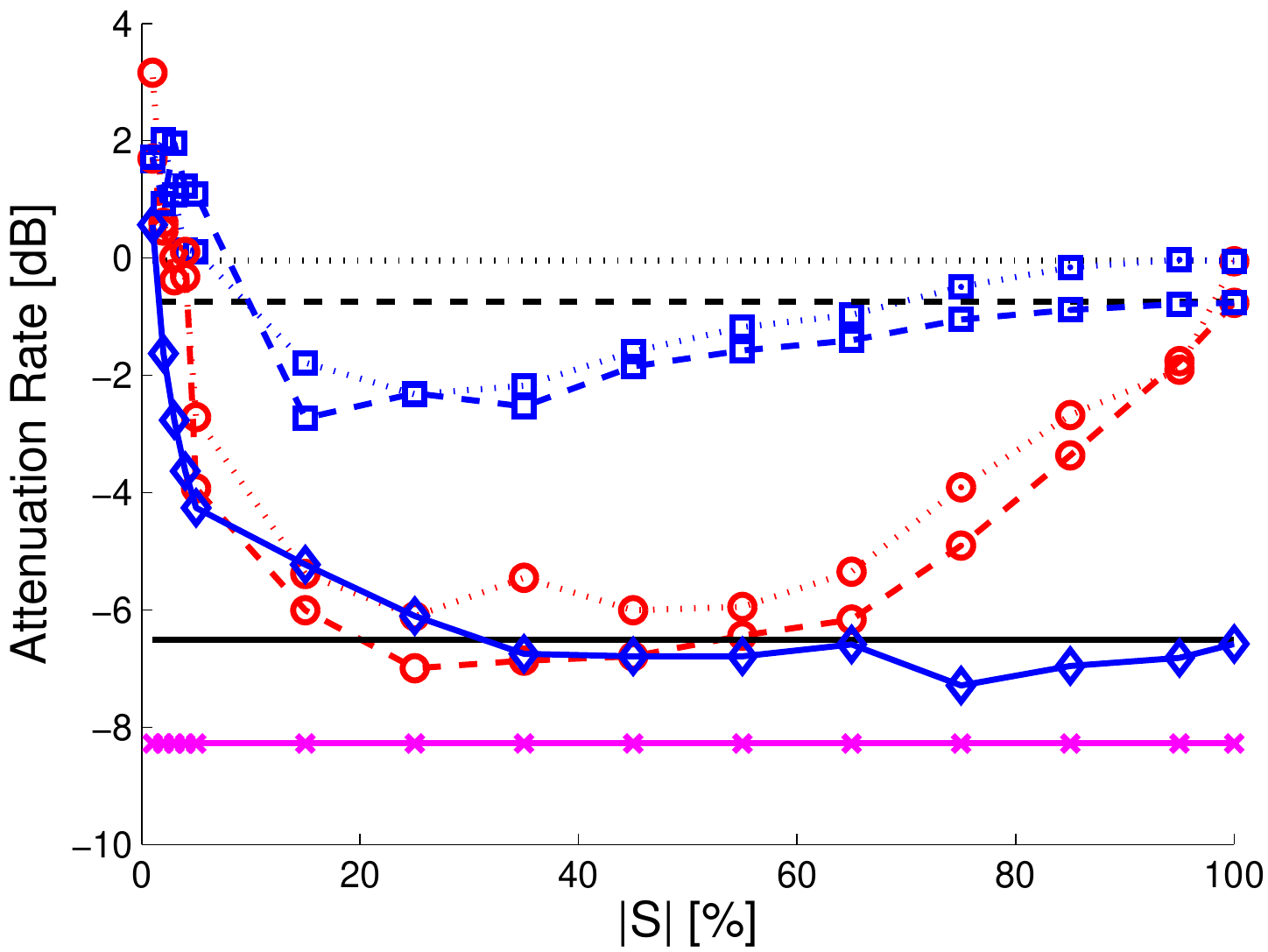}
		\caption{\label{fig:speaker_direct} Female target speaker at $60\degree$ interfered with by a male speaker from the angle of $-60\degree$, both at the distance of $1$~m.}
\end{figure}

A more difficult situation occurs when the interference is a speech signal. We demonstrate this in an experiment where a male speech (interferer) impinges the microphones from the direction of $-60\degree$, while a female speaker (target loudspeaker) is placed at $60\degree$; both at a distance of $1$~m; T$_{60}$ is $160$~ms. The results are shown in Fig.~\ref{fig:speaker_direct}.

Compared to previous experiments, the interfering signal here has similar dynamics and kurtosis as the target signal, which violates the prerequisites of NSFD and of the kurtosis-based selection procedure. Neither FD, NSFD nor FD$^\text{kurt}$ and NSFD$^\text{kurt}$ can distinguish the target speaker from the interfering one and, therefore, all of them perform much worse than FD$^\text{or}$ and NSFD$^\text{or}$ (for a large range of percentage values).

By looking closer at FD and NSFD, they actually try to attenuate both signals by eliminating the dominating signal within each frequency. To show this fact, we performed a simple experiment by taking only the first trial interval of this experiment. Here, FD$^\text{or}$ and NSFD$^\text{or}$ achieved, respectively, attenuation rates of $-7.0$ and $-7.42$~dB with a percentage of 25\%. When the roles of the target and interfering speaker were interchanged so that the oracle procedures took 25\% of frequencies where the interferer was dominant, FD$^\text{or}$ and NSFD$^\text{or}$ attenuated the interferer, respectively, by $11.3$ and $11.2$~dB. The fact that both results were obtained from the same RTF estimates by just selecting different frequency bins confirms that FD and NSFD tend to attenuate both signals. 

In this experiment, we further consider GSS which is capable of blindly separating the target signal from the interference and vice versa\footnote{We apply GSS using known DOAs in this experiment.}. The RTF estimate can be obtained as described in Section~\ref{GSSsection}. Then we can also apply the proposed method based on the selection procedures (\ref{divergencechoice}) and (\ref{coherencechoice}). 

The results in Fig.~\ref{fig:speaker_direct} show that GSS outperforms NSFD as well as FD. Next, GSS$^\text{div}$ (here with $Q=30$)  attenuates the target by about $8$~dB, which improves GSS by $2$~dB. Here GSS$^\text{coh}$ also improves the attenuation rate achieved by GSS, where the best improvement is achieved for 70--80\%. Hence, GSS$^\text{div}$ appears to be better than GSS$^\text{coh}$. However, other experiments not shown here due to space limitations prove that this comparison does not hold in general.

\subsection{Attenuation Rate versus Length of Data}
\label{sec:expLength}
\begin{figure}
	\centering
		\includegraphics[width=0.5\linewidth]{legend_noGSS.pdf}\\
		\includegraphics[width=0.49\linewidth]{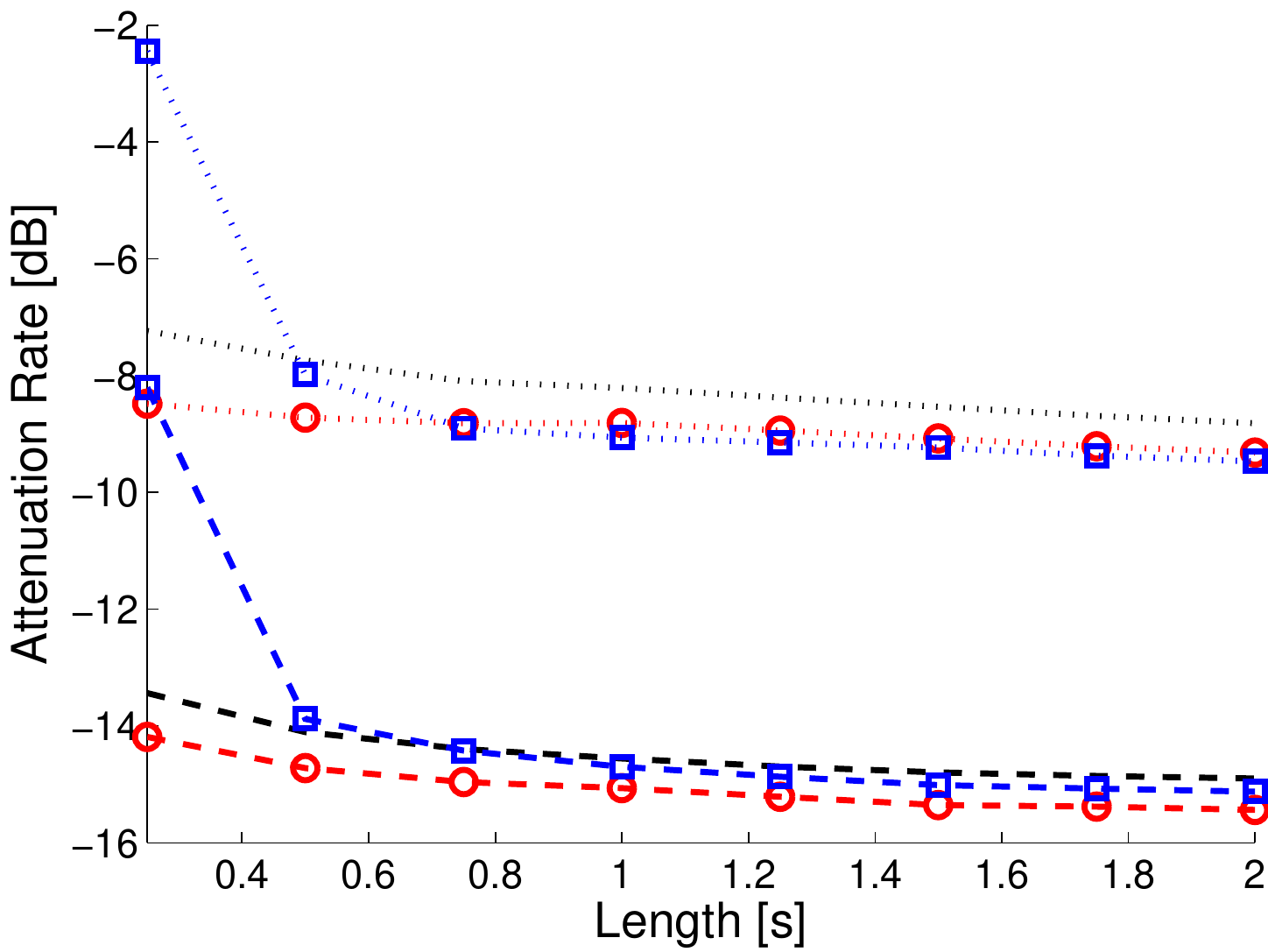}
		\includegraphics[width=0.49\linewidth]{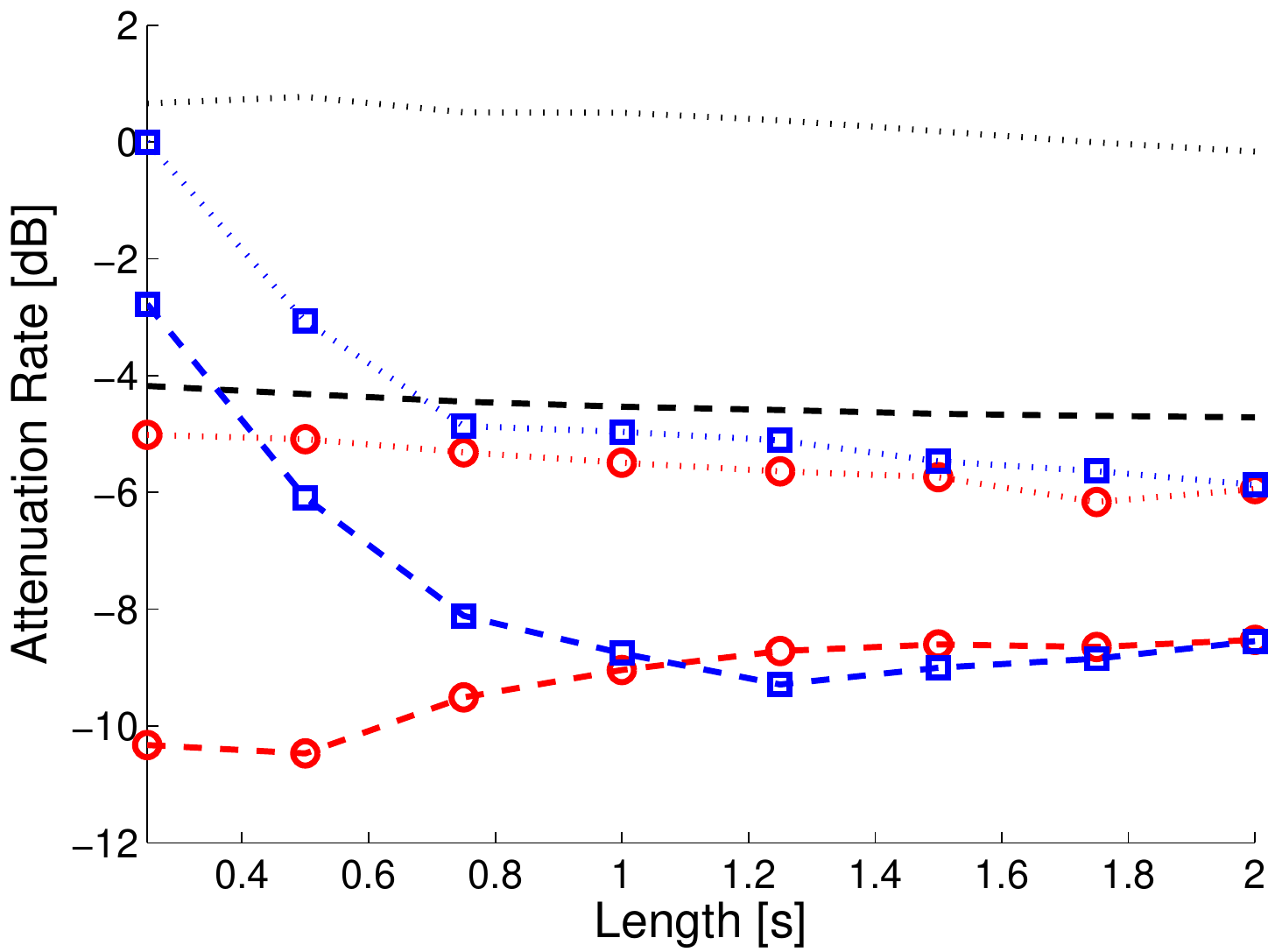}\\
		(a)\hspace{0.5\linewidth}(b)
		\caption{\label{fig:length_direct} Dependence of attenuation rate on the length of data interval. The target speaker is interfered with by (a) temporally and spatially white Gaussian noise and (b) omnidirectional babble noise.}
\end{figure}

Fig.~\ref{fig:length_direct} shows results of repeated experiments, respectively, with temporally and spatially white Gaussian noise and omnidirectional babble noise. The selection percentage of the proposed methods was, respectively, fixed at 25\% and 45\% while the data length was varied from $250$~ms to $2$~s. 

The attenuation rates of FD and NSFD are slowly improved with a growing interval length. Also the performance of the proposed variants is improved with a growing length of data. On the other hand, the improvement is not necessarily monotonic, since the attenuation rate also depends on the percentage, which is fixed in this experiment. An example of the non-monotonic performance is that of NSFD$^{\rm or}$ in Fig.~\ref{fig:length_direct}(b). Next, NSFD$^{\rm kurt}$ and FD$^{\rm kurt}$ perform even worse than NSFD and FD for the data length of $250$~ms. This may be solved by increasing the percentage in the latter methods closer to 100\%. The performances of NSFD$^{\rm or}$ and FD$^{\rm or}$ remain stable for all data lengths, which points to room for possible improvements (e.g. more robust selection procedures).

\subsection{Varying SNR$_\text{in}$}
\begin{figure}
	\centering
	\includegraphics[width=0.5\linewidth]{legend.pdf}\\
		\includegraphics[width=0.49\linewidth]{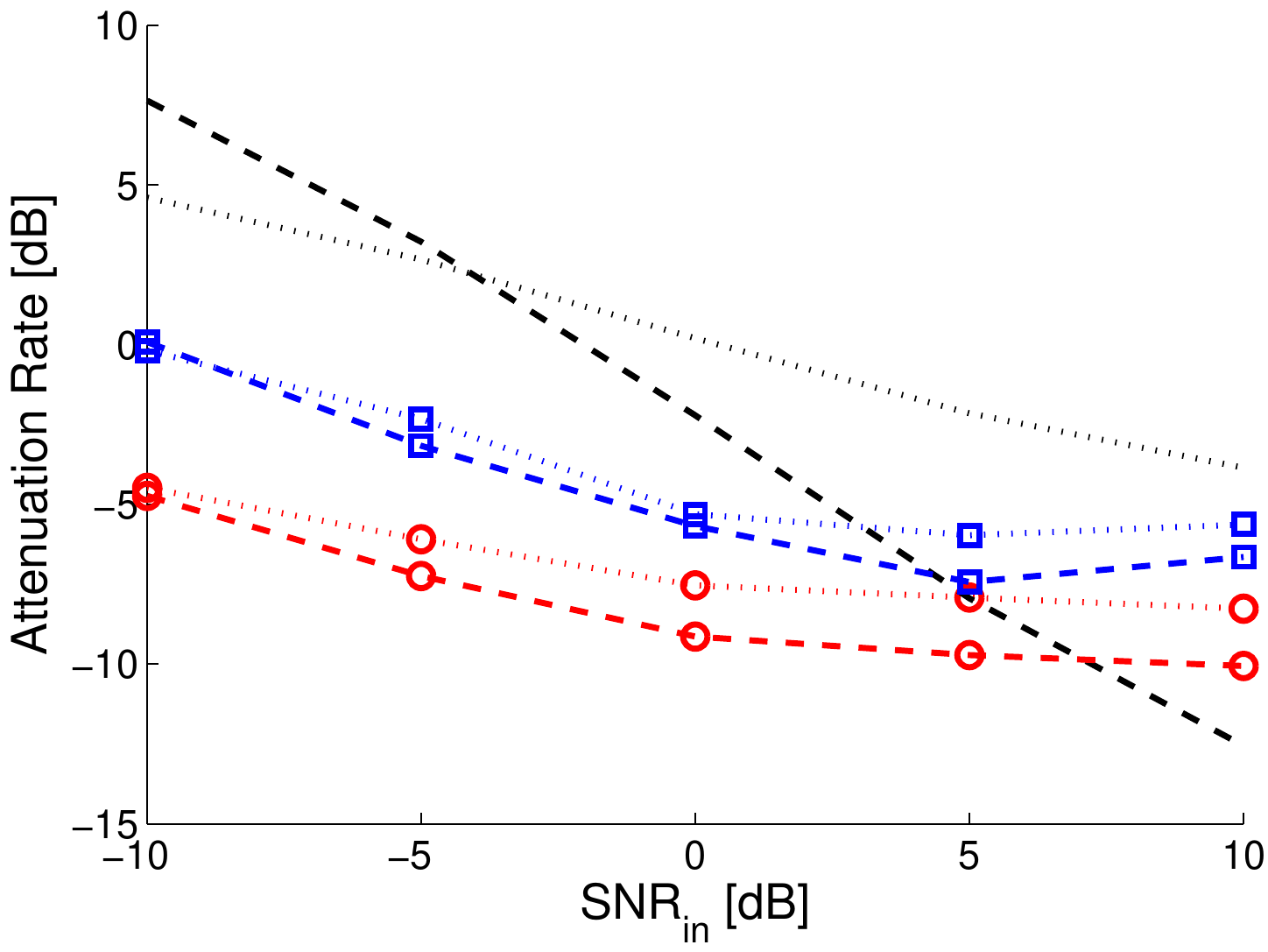}
		\includegraphics[width=0.49\linewidth]{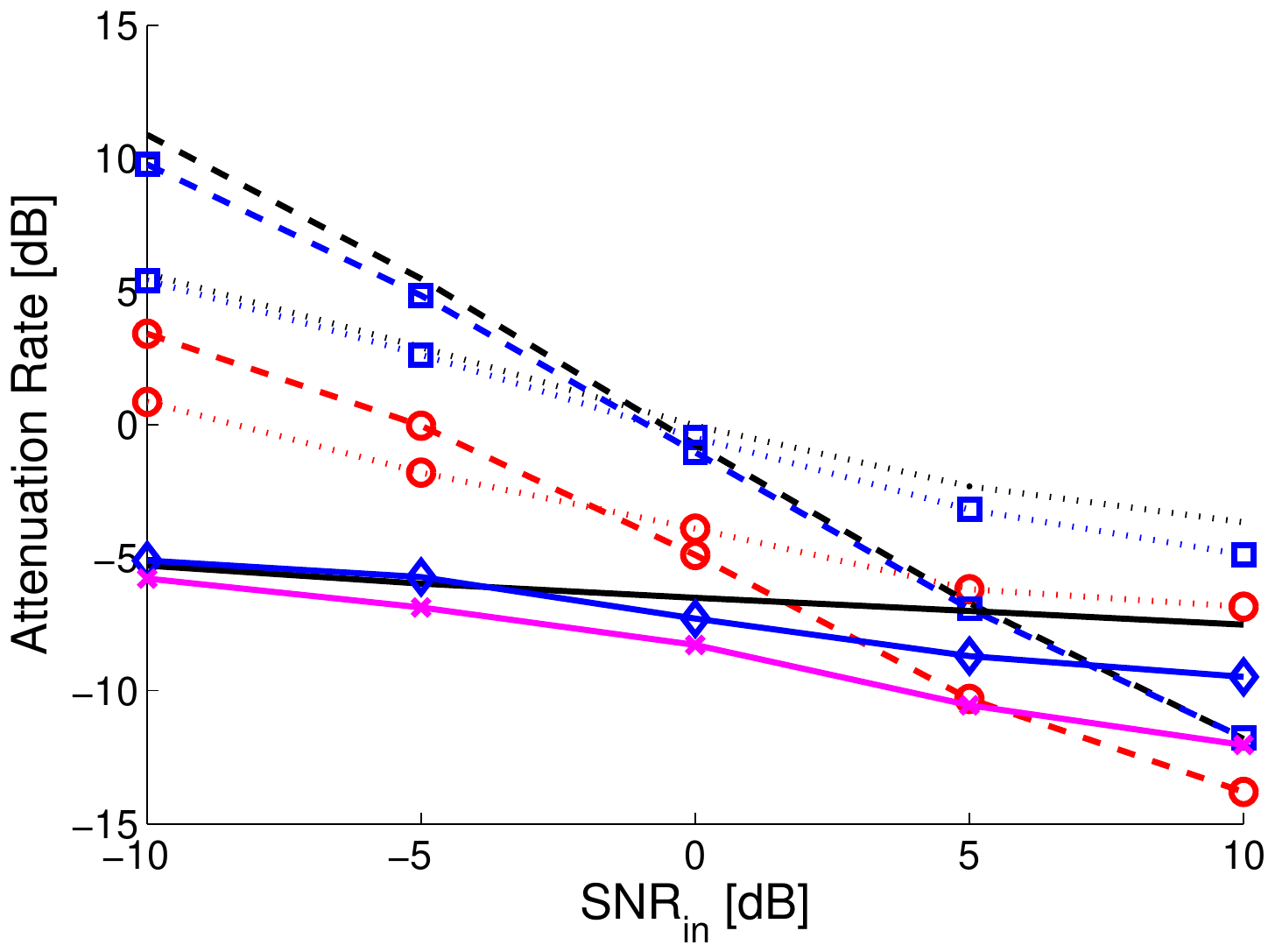}\\
		(a)\hspace{0.5\linewidth}(b)
		\caption{\label{fig:SNR_direct} Attenuation rate as a function of SNR$_\text{in}$ when the target's angle is $60\degree$ and the noise is (a) directional babble coming from a $0\degree$ angle and (b) male speech coming from a $-60\degree$ angle.}
\end{figure}

Here, the experiments where the babble noise was played from a loudspeaker (Fig.~\ref{fig:noise_direct}(b)) and with the male interferer (Fig.~\ref{fig:speaker_direct}) are, respectively, repeated with the percentage fixed, respectively, at 45\% and 55\%; SNR$_\text{in}$ was changed from $-10$ to $10$~dB. Fig.~\ref{fig:SNR_direct} shows the resulting attenuation rates.

The performance of FD and NSFD is improving with growing SNR$_\text{in}$. For SNR$_\text{in}$ below about 0~dB, their attenuation rate goes above zero, because the interfering source is becoming dominant, and FD and NSFD aim to attenuate the former more than the target signal. 

The proposed methods achieve a better attenuation rate than FD and NSFD for almost all values of SNR$_\text{in}$. An exception occurs when SNR$_\text{in}=10$~dB. Here, 
NSFD$^{\rm kurt}$ (and also NSFD$^{\rm or}$ in Fig.~\ref{fig:SNR_direct}(a)) perform worse than NSFD. This is again due to the fixed percentage value, which should be chosen close to 100\% when SNR$_\text{in}$ is high. For SNR$_\text{in}=10$~dB, NSFD appears to be efficient.

In the experiment of Fig.~\ref{fig:SNR_direct}(b), GSS and the variants derived therefrom perform almost constantly and are only slightly improved with the growing SNR$_\text{in}$. This is due to the blind separation of the sources by GSS, which is very efficient when sources are closer to microphones ($1$~m here) and the reverberation time is low (T$_{60}=160$~ms).

\subsection{Varying T$_{60}$}
The last experiment considers varying reverberation time when T$_{60}$ is respectively $160$, $360$ and $640$~ms (the values available in the database \cite{irdatab}). The experiment with two speakers is repeated here with the percentage fixed at 55\%.

\begin{figure}
	\centering
	\includegraphics[width=0.5\linewidth]{legend.pdf}
		\includegraphics[width=0.9\linewidth]{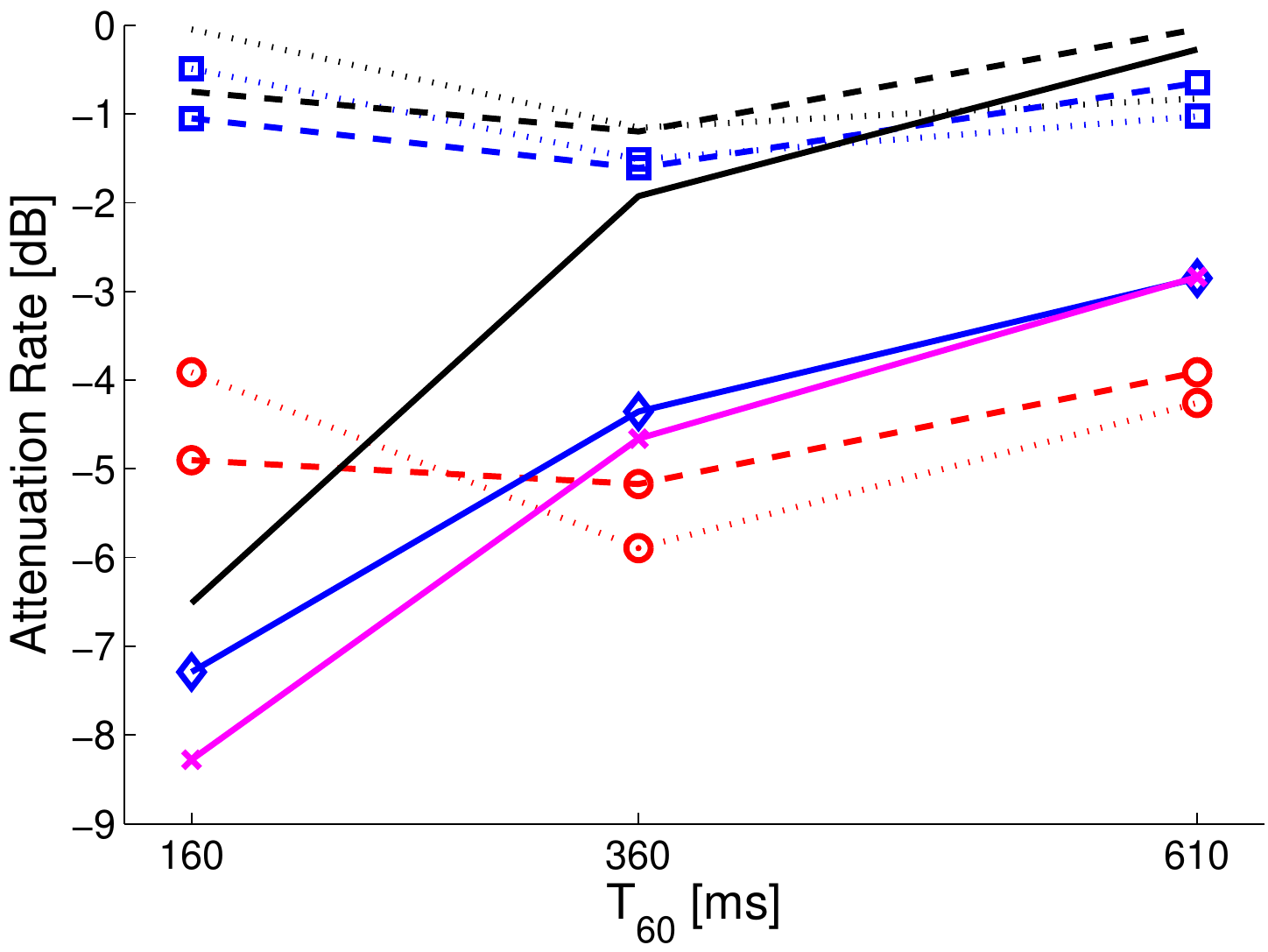}
		\caption{\label{fig:reverb} Attenuation rates as functions of reverberation time $T_{60}$. Female target voice at $60\degree$ was interfered with  a male voice played from the angle of $-60\degree$ both at the distance of $1$~m; SNR$_\text{in}=0$~dB.}
\end{figure}

FS, NSFD and their kurtosis-based variants do not succeed here for any value of $T_{60}$ for the same reasons as in the experiment of Section~\ref{speakersexperiment}. By contrast, the attenuation rates of NSFD$^{\rm or}$ and FD$^{\rm or}$ are only slightly dependent on 
$T_{60}$, which points to the necessity to distinguish the target's and interferer's frequencies correctly. The performance of FD$^{\rm or}$ is even improving with $T_{60}$, but this is again due to the fixed percentage whose optimum value is different for each situation.

The attenuation rate by GSS, GSS$^\text{div}$ and GSS$^\text{coh}$ is dropping as the $T_{60}$ is growing, because the blind separation is becoming difficult with the reverberation time of the environment. Nevertheless, both 
GSS$^\text{div}$ and GSS$^\text{coh}$ improve the attenuation rate by GSS up to by $3$~dB even in the most difficult case when $T_{60}=640$~ms.

\section{Conclusions and Discussion}
We have proposed a novel approach estimating the RTF from noisy data. 
The experiments have shown that, in most situations, the proposed approach yields RTF estimates better than conventional estimators in terms of the capability to cancel the target signal. The crucial parameter to select is the percentage. The optimum percentage depends on many circumstances and is hard to predict. Nevertheless, the experiments where the percentage was fixed have shown that the performance of the method is not too sensitive to this parameter. The performance gain due to the method remains positive when reasonable percentage is chosen, e.g., based on practice.

The proposed method is flexible in providing room for future modifications and improvements, some of which we list now. 

Methods for solving particular parts of the method can be replaced by novel ones, especially the conventional estimators used within the first part. The methods could be tailored to particular scenarios, signal mixtures or noise conditions. For example, we have demonstrated through experiments that NSFD is effective for the first part when noise is isotropic and less dynamic than the target speech signal, while GSS can be efficient when noise is a competitive speech signal.

If some prior knowledge of SNR (or other knowledge) is available, the selection of frequencies (the second part) could be done before or simultaneously with the RTF estimation (the first part). This could lead to computational savings as only the incomplete RTF estimate needs to be computed.

In the method proposed here, the RTF estimate is reconstructed through searching for the sparsest representation of the incomplete RTF in the discrete time-domain. Besides the fact that faster methods for solving (\ref{WLASSO}) may appear in the future, the weighted $\ell_1$ program is by far not the only way to reconstruct the RTF estimate \cite{atomic}. For example, it is possible to reconstruct the RTF in an over-sampled discrete time-domain or in the continuous time-domain; see \cite{atomic2,eusipco2015}. 

Online or batch-online implementations of the proposed methods can be the subject of future developments. For each part, it is possible to select an appropriate online or adaptive method to solve the corresponding task.

\begin{biography}[{\includegraphics[width=1in,height=1.25in,clip,keepaspectratio]{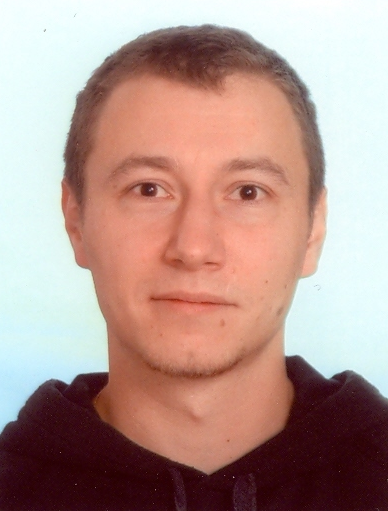}}]{Zbyn\v{e}k Koldovsk\'{y}}
(S'03-M'04) was born in Jablonec nad Nisou, Czech Republic, in 1979. He received the M.S. degree and Ph.D. degree in mathematical modeling from Faculty of Nuclear Sciences and Physical Engineering at the Czech Technical
University in Prague in 2002 and 2006, respectively.

He is currently an associate professor at the Institute of Information Technology and Electronics, Technical University of Liberec. He has also been with the Institute of Information Theory and Automation of the Academy of Sciences of the Czech Republic since 2002. His main research interests are focused on audio signal processing, blind source separation, statistical signal processing, compressed sensing, and multilinear algebra.

Dr. Koldovský serves as a reviewer for several journals such as the IEEE
Transaction on Audio, Speech, and Language Processing, IEEE
Transaction on Signal Processing, Elsevier Signal Processing Journal, and in several conferences and workshops in the field of (acoustic) signal processing. He has served as a co-chair in the fourth community-based Signal Separation Evaluation Campaign (SiSEC 2013) and as the general co-chair of the twelfth International Conference on Latent Variable Analysis and Signal Separation (LVA/ICA 2015).
\end{biography}

\begin{biography}[{\includegraphics[width=1in,height=1.25in,clip,keepaspectratio]{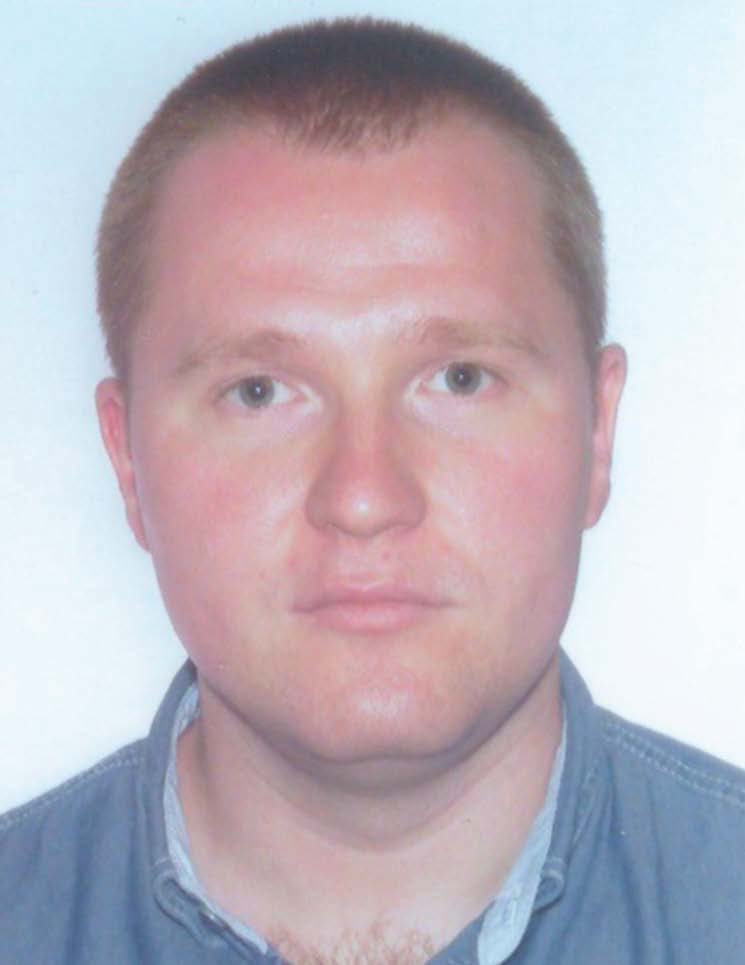}}]{Ji\v{r}\'i M\'alek} received his master and Ph.D. degrees from Technical University in Liberec (TUL, Czech Republic) in 2006 and 2011, respectively,  in technical cybernetics. Currently, he holds a postdoctoral position at the Institute of Information Technology and Electronics, TUL. His research interests include blind source separation and speech enhancement.
\end{biography}

\begin{biography}[{\includegraphics[width=1.25in,height=1.25in,clip,keepaspectratio]{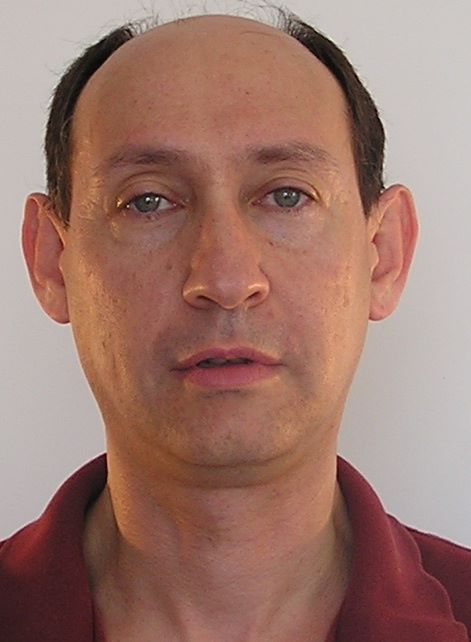}}]
{Sharon Gannot}(S'92-M'01-SM'06) received his B.Sc. degree (summa cum laude) from the Technion Israel Institute of Technology, Haifa, Israel in 1986 and the M.Sc. (cum laude) and Ph.D. degrees from Tel-Aviv University, Israel in 1995 and 2000 respectively, all in electrical engineering. In 2001 he held a post-doctoral position at the department of Electrical Engineering (ESAT-SISTA) at K.U.Leuven, Belgium. From 2002 to 2003 he held a research and teaching position at the Faculty of Electrical Engineering, Technion-Israel Institute of Technology, Haifa, Israel. Currently, he is an Associate Professor at the Faculty of Engineering, Bar-Ilan University, Israel, where he is heading the Speech and Signal Processing laboratory. Prof. Gannot is the recipient of Bar-Ilan University outstanding lecturer award for 2010 and 2014.

Prof. Gannot has served as an Associate Editor of the EURASIP Journal of Advances in Signal Processing in 2003-2012, and as an Editor of two special issues on Multi-microphone Speech Processing of the same journal. He has also served as a guest editor of ELSEVIER Speech Communication and Signal Processing journals.
Prof. Gannot has served as an Associate Editor of IEEE Transactions on Speech, Audio and Language Processing in 2009-2013. Currently, he is a Senior Area Chair of the same journal. He also serves as a reviewer of many IEEE journals and conferences.
Prof. Gannot is a member of the Audio and Acoustic Signal Processing (AASP) technical committee of the IEEE since Jan., 2010. He is also a member of the Technical and Steering committee of the International Workshop on Acoustic Signal Enhancement (IWAENC) since 2005 and was the general co-chair of IWAENC held at Tel-Aviv, Israel in August 2010. Prof. Gannot has served as the general co-chair of the IEEE Workshop on Applications of Signal Processing to Audio and Acoustics (WASPAA) in October 2013. Prof. Gannot was selected (with colleagues) to present a tutorial sessions in ICASSP 2012, EUSIPCO 2012, ICASSP 2013 and EUSIPCO 2013.  Prof. Gannot research interests include multi-microphone speech processing and specifically distributed algorithms for ad hoc microphone arrays for noise reduction and speaker separation; dereverberation; single microphone speech enhancement and speaker localization and tracking.
\end{biography}

\end{document}